\documentclass[%
 aip,
 amsmath,amssymb,
preprint,%
,floatfix]{revtex4-1}

\usepackage{graphicx}% Include figure files
\usepackage{dcolumn}% Align table columns on decimal point
\usepackage{bm}% bold math
\usepackage[utf8]{inputenc}
\usepackage[T1]{fontenc}
\usepackage{mathptmx}
\usepackage{etoolbox}

\usepackage[locale=US,per-mode=reciprocal]{siunitx}
\usepackage[dvipsnames]{xcolor}
\usepackage{soul}
\usepackage[ruled,linesnumbered]{algorithm2e}

\SetCommentSty{mycommfont}
\usepackage{algpseudocode}
\usepackage{chemformula}
\usepackage{subcaption}
\usepackage{amsmath}
\usepackage{physics}
\usepackage{tikz}
\usetikzlibrary{shapes, arrows}

\tikzstyle{startstop} = [rectangle, rounded corners, minimum width=3cm, minimum height=1cm, draw=black, text centered]
\tikzstyle{process} = [rectangle, minimum width=3cm, minimum height=1cm, draw=black, text centered]
\tikzstyle{arrow} = [thick,->,>=stealth]

\newcommand{\mm}[1]{\mathrm{#1}}
\newcommand{\bb}[1]{\mathrm{\textbf{#1}}}
\newcommand*\diff{\mathop{}\!\mathrm{d}}

\newcommand{\kb}{k_\mm{B}}

\makeatletter
\def\@email#1#2{%
 \endgroup
 \patchcmd{\titleblock@produce}
  {\frontmatter@RRAPformat}
  {\frontmatter@RRAPformat{\produce@RRAP{*#1\href{mailto:#2}{#2}}}\frontmatter@RRAPformat}
  {}{}
}%
\makeatother
\begin{document}

% \preprint{AIP/123-QED}

\title[Chemical Reactions in Rarefied Gas Flows]{Modeling of Chemical Reactions in Rarefied Gas Flows by the Kinetic Fokker-Planck Method}\thanks{The following article has been submitted to/accepted by Physics of Fluids. After it is published, it will be found at https://pubs.aip.org/aip/pof.}
% Force line breaks with \\
\author{L. Basov}
\affiliation{ 
German Aerospace Center (DLR), Goettingen, Germany.%\\This line break forced with \textbackslash\textbackslash
}%
\affiliation{ 
Applied and Computational Mathematics, RWTH Aachen, Aachen, Germany.%\\This line break forced with \textbackslash\textbackslash
}%
\author{G. Oblapenko}
\affiliation{ 
Applied and Computational Mathematics, RWTH Aachen, Aachen, Germany.%\\This line break forced with \textbackslash\textbackslash
}%
\author{M. Grabe}%
 \email{leo.basov@dlr.de}
\affiliation{ 
German Aerospace Center (DLR), Goettingen, Germany.%\\This line break forced with \textbackslash\textbackslash
}%

% \date{\today}% It is always \today, today,
%              %  but any date may be explicitly specified

\begin{abstract}
We propose a novel approach for modeling chemical reactions within the particle-based Fokker-Planck framework for gas flow simulations which conserves mass, momentum, and energy while retaining the performance advantages of the Fokker-Planck approach over the Direct Simulation Monte Carlo (DSMC) method in areas of high density.
We show an application of the approach to recombination and exchange reactions, discuss verification results, and demonstrate performance advantages when compared to DSMC for applications in low Knudsen number regimes.
The developed method can be applied to simulation of flows in the continuum and transitional regimes, as well as to multi-scale coupled Fokker-Planck-DSMC simulations.
\end{abstract}

\maketitle

\section{Introduction}
Flows encountered in aerospace applications, like atmospheric entry or thruster plume expansion and interactions, can be described by the well-known Boltzmann equation.
A common approach to solve it numerically is the Direct Simulation Monte-Carlo (DSMC) method pioneered by Bird \cite{bird_molecular_1994}.
The method is very efficient for moderate to high Knudsen numbers but becomes increasingly computationally intensive when approaching the continuum limit.
In this regime the Boltzmann equation can be numerically solved using the kinetic Fokker-Planck (FP) method \cite{jenny_solution_2010} which, like DSMC, relies on simulated particles to model the transport of mass, momentum and energy through the flow domain, but unlike DSMC does not resolve individual collisions.
Due to the similarity in their formulation a hybrid method based on DSMC and FP can be derived \cite{gorji_fokkerplanckdsmc_2015}.
Unlike the coupling of Navier-Stokes type solvers to DSMC~\cite{hash1997two,roveda1998hybrid}, the FP-DSMC method does not encounter problems on model interfaces as the hybrid method only relies on switching between the velocity updates. Problems with the treatment of boundary conditions are also not present as those are identical between DSMC and FP.
The hybrid method provides a speedup of up to $\sim5$ compared to pure DSMC simulations without notable loss of accuracy even for complex simulation setups \cite{kuchlin_parallel_2017, kuchlin_automatic_2018, jun_assessment_2018, jun_cubic_2019} which makes the approach a powerful tool to investigate flows on a wide range of Kn  numbers.
In the last years, the FP model has been extended to describe diatomic \cite{gorji_fokkerplanck_2013,hepp_master_2020} and polyatomic \cite{mathiaud_fokkerplanck_2017, aaron_nagel_modeling_2023, basov_modeling_2024} gases for single species  as well as for mixtures \cite{hepp_kinetic_2020,hepp_kinetic_2020-2, hepp_kinetic_2022}.
However, the modeling of chemical reactions in the kinetic FP method remains an open research topic.

%High performance implementation and analysis done by Kuchlin\cite{kuchlin_parallel_2017}.
%He extensively investigated the hybrid approach also in the context of grid refinement  \cite{kuchlin_automatic_2018, kuchlin_stochastic_2018}.

%Jun assessed the hybrid FP-DSMC approach (grid resolution, adaptation, performance and accuracy compared to DSMC) for monoatomic \cite{jun_assessment_2018} and diatomic \cite{jun_cubic_2019} cubic FP.

%\hl{Below is a part discussing modelling of chemistry in non-equilibrium flows in general. Need to somehow make it fit better with the other parts in the introduction. Perhaps put the FP stuff afterwards, saying that basically a lot of the methods (Chapman-Enskog, DSMC, etc.) have considered chemical reactions, but in FP it's basically an open topic without prior work.}
In rarefied regimes, the distributions of the particles' velocities and internal energies can deviate from the equilibrium Maxwell-Boltzmann statistics, which in turn can affect reaction rates, and needs to be accounted for in flow simulations. First theoretical works in the field considered the impact of non-equilibrium velocity distributions on chemical reaction rates~\cite{prigogine1949perturbation,takayanagi1951theory,shizgal1970nonequilibrium} via the Chapman--Enskog method; this impact was also observed in DSMC simulations~\cite{lemarchand1998perturbation}. More recent results for reacting flows with non-equilibrium velocity distributions include the proof of the existence of a stable equilibrium solution of the Boltzmann equation for reacting flows~\cite{rossani1999note}, and development of moment-based approaches~\cite{sarna2021moment}. The approach based on the Chapman--Enskog expansion has been also extended to polyatomic gas flows in thermal equilibrium~\cite{kustova2011cross} and non-equilibrium~\cite{kolesnichenko2010gas,kustova2015reaction}.

In the Direct Simulation Monte Carlo method, the most widely used approach to simulation of chemical reactions remains the Total Collision Energy (TCE) model~\cite{bird_molecular_1994,bird_chemical_2011}, which aims to reproduce the Arrhenius law in the equilibrium case. However, the model is not necessarily valid outside of equilibrium; it also places restrictions on the constant appearing in the Arrhenius law (thus, not every Arrhenius curve fit can be represented in TCE) and the internal energies (which are assumed to be continuous).
Various modifications to the model have been proposed~\cite{wadsworth1997vibrational,gimelshein_use_2004} to address these issues, however, the standard formulation remains most widely-used due to its simplicity.
Recent simulations coupling DSMC and molecular dynamics simulations have confirmed the role of coupling between internal energy relaxation and chemical reactions~\cite{grover2019direct}, thus again highlighting the need for chemical reaction models that account for non-equilibrium effects.

Approaches to simulation of chemically reacting flows based on the model Bhatnagar--Gross-Krook (BGK) equation have so far been applied only to continuum flows~\cite{groppi2004bhatnagar,monaco2005bgk,baranger2020bgk}; moreover, due to the use of the BGK approximation, they assume an equilibrium velocity distribution.

%Chemistry - here I need more litrature
%Many problems arise from the fact that properties of the reactants need to be considered to evaluate a chemical reaction.
%This contradicts the FP approach that achieves its linear scaling in terms of particle numbers by forgoing the evaluation of interaction pairs.
In this paper we propose a novel approach to modeling chemical reactions for an arbitrary number of constituents via our cubic FP method implemented in the SPARTA framework \cite{plimpton_direct_2019}.
The model conserves mass, energy and momentum while retaining the performance advantages in areas of low Kn numbers.
It is verified by comparing to continuum-based analytical and numerical solutions.
The main advantage of the presented approach is that our model can work with arbitrary reaction rates.
Those can be formulated in a such a way as to consider the internal energy distribution of the reactants, therefore providing better modeling of thermal non-equilibrium compared to BGK based approaches.
However, the first implemented of the model used in this paper to showcase and verify the novel approach uses reaction rates which are based on equilibrium assumption without loss of generality.
The implementation and development of reaction rates considering internal energy modes is left for future work.
%Since the goal of the FP model is to efficiently simulate gas flows in the continuum limit, the role of the aforementioned non-equilibrium effects is expected to be less significant.

The paper is structured as follows: first, an overview of the Fokker-Planck method is given.
Its extension to chemically reacting flows is then proposed. Comparisons to analytical and numerical solutions are then presented for various testcases.
Finally, an analysis of the computational performance of the model in comparison to DSMC is presented.
\section{Method}
In this section we briefly review the general idea of the kinetic Fokker-Planck method and point out the challenges one faces when extending the model to chemical reactions.
We describe the newly proposed chemistry algorithm and and provide  details on its workings and its ability to conserve lower moments of the velocity distribution function. 
We exemplary apply the proposed approach to model dissociation and exchange reactions.

\subsection{Cubic Fokker-Planck Method}
We base our further discussion on the cubic formulation of the Fokker-Planck method, which is covered in detail in the literature \cite{gorji_fokkerplanck_2011, jun_assessment_2018, hepp_master_2020} and thus will be only briefly summarized here.
The implementation of chemistry requires additionally a formulation of mixtures in the FP framework which was derived for hard sphere (HS) \cite{hepp_kinetic_2020} as well as variable hard sphere (VHS) interaction potentials \cite{hepp_kinetic_2020-1}.
Even though the chemistry algorithm presented in the present work was implemented using the cubic FP method, it can be applied to other formulations, like the entropic FP\cite{gorji_entropic_2021}, as well.

The well-known Boltzmann equation describes the evolution of a scalar distribution function $f$ due to intermolecular collisions in rarefied flows:
%----------------------------------------------------------------------------------
\begin{equation}
    \label{eqn:boltzmann}
    \frac{\mathrm{D}f}{\mathrm{D}t} = S_\mathrm{Boltz}\;,
\end{equation}
%----------------------------------------------------------------------------------
where $t$ is time and $S_\mathrm{Boltz}$ is the Boltzmann collision integral.
%----------------------------------------------------------------------------------
One approach to reduce the computational complexity of the integro-differential Eqn.~(\ref{eqn:boltzmann}) is to approximate $S_\mathrm{Boltz}$ by a Fokker-Planck collision operator~\cite{jenny_solution_2010}, $S_\mathrm{FP}$:
%----------------------------------------------------------------------------------
\begin{equation}
    \label{eqn:fokker_planck}
    S_\mathrm{Boltz} \approx S_\mathrm{FP} = -\frac{\partial}{\partial V_i} (A_i f) + \frac{1}{2}\frac{\partial^2}{\partial V_j \partial V_j} (D^2 f)\;,
\end{equation}
%----------------------------------------------------------------------------------
where the drift coefficient $A_i$ and the diffusion coefficient $D$ of Eqn.~(\ref{eqn:fokker_planck}) are model parameters chosen in such a way that production terms calculated using the Boltzmann collision operator are reproduced in the continuum limit.
According to It\^{o}'s lemma, Eqn.~(\ref{eqn:fokker_planck}) can be written as a set of stochastic differential equations
%----------------------------------------------------------------------------------
\begin{align}
    \label{eqn:dx_dt}\frac{\mm{d}\bb{X}}{\mm{d}t} &= \bb{V}, \\
    \label{eqn:dv_dt}\frac{\mm{d}\bb{V}}{\mm{d}t} &= \bb{A} + D \frac{\mm{d}\bb{W}}{\mm{d}t} + \bb{F},
\end{align}
%----------------------------------------------------------------------------------
with $\bb{X}$ denoting particle position, $\bb{V}$ the particle velocity and $\mathrm{d}\bb{W}$ the Wiener process with zero expectation. Thus, the resulting Fokker-Planck equation can be solved through stochastic motion which in turn can be modeled using a particle method.

The choice of the order of the reproduced production terms defines the structure of the chosen model parameters.
For the cubic approach productions terms up to the heat flux are recovered in the continuum limit.
Using the definition for thermal velocity $\bb{M} = \bb{V} - \bb{U}$, where $\bb{U}$ is the bulk or center of mass velocity of the gas for a single species gas, the drift $A_i$ and the diffusion coefficient $D$ are chosen as
%----------------------------------------------------------------------------------
\begin{align}
    A_i &= -\frac{1}{\tau} M_i + \xi_{ij} M_j + \gamma_i \left( M_j M_j - \frac{3 k_\mathrm{B} T}{m} \right) + \Lambda \left(M_i M_j M_j - \frac{2 q_i}{\varrho} \right), \\
    D &= \sqrt{\frac{2 k_\mathrm{B} T}{\tau m}}\;.
\end{align}
%----------------------------------------------------------------------------------
where $\tau = 2 \mu / p$ is the relaxation time, $\gamma$ and $\xi$ are the coefficient matrices, $k_\mathrm{B}$ is the Boltzmann constant, $T$ the temperature, $m$ is the mass, $\bb{q}$ the heat flux vector and $\varrho$ the (mass) density. 
The parameter $\Lambda$ is chosen to assure stability of the model.
Integrating equation~(\ref{eqn:dv_dt}) leads to the equation describing the motion of the particles in the form
%----------------------------------------------------------------------------------
\begin{multline}
    V^{n+1} = \frac{1}{\alpha} \left[ M_i^n \exp(-\frac{\Delta t}{\tau}) + \left(1 - \exp(-\frac{\Delta t}{\tau}) \right)  \tau N^n_i \right.\\
    +\left. \psi_i \sqrt{\frac{D^2}{2} \tau \left( 1 - \exp(-2 \frac{\Delta t}{\tau}) \right)} \right] + U_i
\end{multline}
%----------------------------------------------------------------------------------
where the exponent $n$ denotes the n\textit{th} time step and $\psi_i$ are independent standard normal variates.
Here $\bb{N}$ designates the non-linear term
%----------------------------------------------------------------------------------
\begin{equation}
    N_i^n = \xi_{ij} M_j^n + \gamma_i \left(M_j^n M_j^n - \frac{3 k_\mathrm{B} T}{m} \right) + \Lambda \left(M_i^n M_j^n M_j^n - \frac{2 q_i}{\varrho} \right)\;.
\end{equation}
%----------------------------------------------------------------------------------
The factor $\alpha$ is a parameter used to scale particle velocities to ensure energy conservation for arbitrary time steps.
To be compatible with DSMC, a simple integration scheme for particle positions is chosen as
%----------------------------------------------------------------------------------
\begin{equation}
    \bb{X}^{n+1} = \bb{X}^{n} + \bb{V}^{n} \Delta t\;.
\end{equation}
%----------------------------------------------------------------------------------

\subsection{Proposed Algorithm}
%The main advantage of the FP method compared to DSMC lies in the fact that the velocity update for a single particle is performed using only the information about said particle and the velocity moments calculated using the particle ensemble in the cell.
%It means that no direct collision pairs need to be made and evaluated which allows FP to have a constant computational time in terms of the Knudsen number.
%This is not the case for DSMC as the number of collision pairs to be considered rises inversely with the Knudsen number, making FP more efficient in areas of high density (low Knudsen numbers).
%The advantage is somewhat upset by FP under predicting the sharpness of shock structures.
%This can be remedied to some degree by using the entropic formulation of FP \cite{gorji_entropic_2021} or using a hybrid approach \cite{gorji_fokkerplanckdsmc_2015}.
%While the former provided a more rigorous structure and a proof of the H-theorem while the later uses DSMC to correctly resolve areas of sharp gradients in the flow field.

In the past the FP equation has been used to model chemical reactions and was shown to be an adequate approximation of the chemical master equation \cite{grima_how_2011}.
However, to the authors knowledge it has not been applied to model flow problems including chemical reactions by the means of a particle method.

The kinetic FP method achieves its linear scaling in terms of the Knudsen number by resolving an evolution of the individual particle velocities without choosing and evaluating particle interaction pairs as it is done in the DSMC method.
However, this approach poses a problem for the modeling of chemical reactions as no information is available during the reaction step about a potential reaction partner and their properties like momentum or energy.
To solve this problem we propose a novel algorithm that keeps the Knudsen number-independent scaling of FP but extends its modeling capability to chemically reacting flows.

%\subsubsection{Overview}
%The DSMC approach uses particles pairs to evaluate reaction energies and post-collision states, and conserve moments of the velocity distribution function.
%In continuum methods only the state of the (cell) local system is considered when evaluating chemistry.
%In our approach we try to use as much information as possible from available particles during the reaction step as to retain some information of the local distribution function while only using the FP approach of linearly passing though the list of available particles.
%By doing so we use the information of the a particle currently selected in the cell plus the properties calculated from the cell ``background'' particles.
%This poses several challenges.
%For one there is no information available to evaluate the available reaction energy for each particle-background pair in such a way as to exactly conserve energies.
%Additionally, the reaction probability based on the chosen kinetic particle and the background gas needs to be evaluated.
%The question exactly which of the other particles constituting the background during the evaluation of the reaction will participate in the reaction needs to be addressed. 
%An overview of the algorithm as it is executed in each cell is seen in Fig.~\ref{fig:chem_flowchart}.

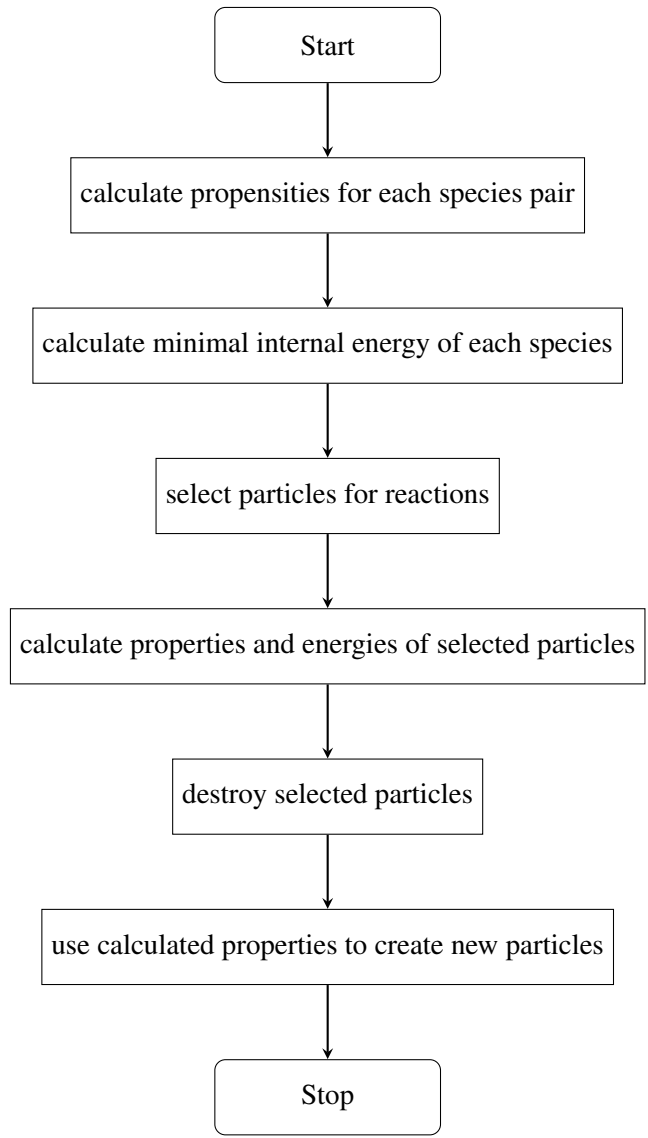
\begin{figure}
  \centering
  \begin{tikzpicture}[node distance=2cm, auto]

    % Nodes
    \node (start) [startstop] {Start};
    \node (calculate) [process, below of=start] {calculate propensities for each species pair};
    \node (process) [process, below of=calculate] {calculate minimal internal energy of each species};
    \node (select) [process, below of=process] {select particles for reactions};
    \node (calc_props) [process, below of=select] {calculate properties and energies of selected particles};
    \node (destroy) [process, below of=calc_props] {destroy selected particles};
    \node (create) [process, below of=destroy] {use calculated properties to create new particles};
    \node (stop) [startstop, below of=create] {Stop};

    % Arrows
    \draw [arrow] (start) -- (calculate);
    \draw [arrow] (calculate) -- (process);
    \draw [arrow] (process) -- (select);
    \draw [arrow] (select) -- (calc_props);
    \draw [arrow] (calc_props) -- (destroy);
    \draw [arrow] (destroy) -- (create);
    \draw [arrow] (create) -- (stop);

  \end{tikzpicture}
  \caption{Chemistry process performed in each cell}
  \label{fig:chem_flowchart}
\end{figure}

The DSMC approach relies on particle pairs to evaluate reaction energies, determine post-collision states, and conserve the moments of the velocity distribution function. In contrast, continuum methods consider only the state of the local (cell-averaged) system when evaluating chemical processes.
Our approach aims to leverage as much information as possible from the available particles during the reaction step to retain some aspects of the local velocity distribution function. This is achieved while adhering to the linear particle processing scheme of the FP method, where particles are processed sequentially. Specifically, we utilize the properties of the currently selected particle in the cell, along with those derived from the cell's "background" particles.
This method presents several challenges. First, there is insufficient information to accurately evaluate the reaction energy for each particle-background pair in a manner that strictly conserves energy. Second, the reaction probability must be calculated based on the interaction between the selected particle and the background gas. Finally, determining which background particle will participate in the reaction poses a nontrivial problem.
A detailed overview of the algorithm as implemented in each cell is illustrated in Fig.~\ref{fig:chem_flowchart}.
%\begin{enumerate}
%    \item Calculate per species values
%    \begin{itemize}
%        \item propensities
%        \item min internal energies
%    \end{itemize}
%    \item iterate over each particles
%    \begin{itemize}
%        \item if species counter > 0 add particle for removal
%        \item else if reaction prob > 0 and potential partner available add to reacted
%        \item else continue
%    \end{itemize}
%    \item use particles marked for reaction to calculated energies and momentum
%    \item use values to create products for reactions
%    \item remove particles marked for removal and particles participated in reactions
%\end{enumerate}

\subsection{Derivation of Reaction Probability}
Our first concern is to evaluate the probability for a reaction to take place for a given particle.
We base our approach on Gillespie's direct simulation method \cite{gillespie_general_1976, gillespie_stochastic_2007}.
A very similar approach was utilized in calculating the transition probabilities of internal energy states using the master equation ansatz to model diatomic \cite{hepp_master_2020} and polyatomic \cite{basov_modeling_2024}  molecules in FP.

Gillespie assumes that the mixture under consideration is well-stirred, confined to a constant volume $\Omega$, and is in thermal (but not chemical) equilibrium at some constant temperature.
The chemical reactions are modeled as instantaneous processes.

In the kinetic FP model, moments of the distribution function are evaluated on a cell-based level under the assumption that gas properties remain constant throughout the cell and do not change during the time step. This assumption aligns with the premise of the mixture being (locally) well-stirred and confined to a constant volume.
Since the FP model is applicable primarily in conditions close to the continuum regime, the assumption of thermal equilibrium is considered valid. Consequently, the temperature is treated as constant over the duration of a given time step.

%The original formulation of Gillespie is also based on the assumption that the system can be described by the molecular population alone ignoring positions and velocities of the molecules.
%The justification lies, according  to Gillespie, in the conditions for the system being well stirred.
%One fundamental assumption made in the argument is that the vast majority of collision that take place are elastic which leads to two effects
%\begin{enumerate}
%	\item the particles become uniformly distributed throughout the system volume
%	\item the velocity distribution of the molecules becomes the Maxwell-Boltzmann distribution
%\end{enumerate}
%Those assumptions are used to simplify the problem by ignoring the modeling of elastic collisions. More specific the modeling of non reactive collisions as elastic collisions that lead to change of internal energies of molecules are not considered as well. 
%\hl{This is an implication that chemistry happens on time scales which are much larger than relaxation times of the gas.}
%The effect that is modeled is the change of population of species exclusively by the means of chemical reactions.
%The chemical reactions are modeled as instantaneous processes.
%This means that when a reaction $j$ occurs the systems immediately jumps from state $\vec{x}$ to $\vec{x} + \vec{\nu}_j$ where $\vec{\nu}_j$ is the sate change vector for the reaction $j$.
%\hl{Say: Total change in population.}

According to Gillespie, the main premise of stochastic chemical kinetics is the definition of the so called propensity function $a_j$ of the reaction $j$.
Namely, $a_j( \vec{\mathfrak{Z}})\diff{t}$ is the probability that given the state of the system, described by vector of particle numbers of individual species $\vec{\mathcal{Z}}(t) = \vec{\mathfrak{Z}}$ at time $t$, one $j$ reaction will occur somewhere inside the system volume in the next infinitesimal time interval $[t, t + \diff{t})$ \cite{gillespie_stochastic_2007}.

The rest of the theory can be derived from the definition of $a_j$ via the laws of probability.
Gillespie provides a physical rationale for the propensity function applying it to unimolecular and bimolecular reactions.
Here we will only focus on the bimolecular ones.
If the reaction $j$ is bimolecular and of the form $\ch{S1 + S2 ->}$ product(s), Gillespie states that there exists a constant $c_j$ such that $c_j \diff{t}$ gives the probability that a randomly chosen pair of \ch{S1} and \ch{S2} molecules will react according to $j$ in the next infinitesimal time $\diff{t}$.
Assuming the number of \ch{S} particles in volume $\Omega$ is $\mathfrak{Z}$, the probability that some of the \ch{S1} - \ch{S2} pairs in the volume will react according to $j$ is $\mathfrak{Z}_1 \mathfrak{Z}_2 c_j \diff{t}$ and the propensity function is
\begin{equation}
	a_j(\vec{\mathfrak{Z}}) = c_j \mathfrak{Z}_1 \mathfrak{Z}_2.
\end{equation}
For a reaction of the type $\ch{S1 + S1 ->}$ product(s) it is accordingly
\begin{equation}
	a_j(\vec{\mathfrak{Z}}) = \frac{1}{2} c_j \mathfrak{Z}_1 (\mathfrak{Z}_1 - 1).
\end{equation}
The constant $c_j$  is numerically equal to
\begin{equation}
c_j = 
\begin{cases}
    \frac{k_j}{\Omega}, & \ch{S1} \neq \ch{S2}\\
    2 \frac{k_j}{\Omega},& \ch{S1} = \ch{S2}
\end{cases}
\end{equation}
where $k_j$ is the reaction-rate constant of conventional deterministic chemical kinetics.

Since we are dealing with simulated particles of which every particle represents a number $w$ of real particles as
\begin{equation}
	\mathfrak{Z}_i = w N_i
\end{equation}
we can write the propensity function for $S_1 \neq S_2$ as
\begin{equation}
	a_j(\vec{\mathfrak{Z}}) = w^2 N_1 N_2 \frac{k_j}{\Omega}
\end{equation}
and
\begin{equation}
	a_j(\vec{\mathfrak{Z}}) = w^2 N_1 (N_1 - 1) \frac{k_j}{\Omega}
\end{equation}
otherwise.
We want to check or ``divide'' the probability between batches of particles, namely particles that fulfill the requirement for the given reaction $j$ to have sufficient energy to pass the test $e_{\mm{int}, k} + \min(e_{\mm{int}, b}) \leq \Delta e$ as defined under energy conservation.
The number of these particles is $\mathfrak{Z}_p = w N_p$ which leads to the propensity functions
\begin{align}
	a_j(\vec{\mathfrak{Z}}) &= w \frac{N_1 N_2}{N_p} \frac{k_j}{\Omega}\\
	a_j(\vec{\mathfrak{Z}}) &= w \frac{N_1 (N_1 - 1)}{\frac{1}{2}N_p} \frac{k_j}{\Omega}.
\end{align}
%\hl{Here I need a justification for the factor $\frac{1}{2}$.
%I think it comes from he fact that $x_p$ for equal species must be $\frac{1}{2} x_p$ as coming from the "symmetry factor".}

The definition of the propensity function can be used to derive a probability to model trajectories of $\vec{\mathcal{Z}}(t)$.
This is not the same as solving chemical master equation numerically, which would give the probability density function of $\vec{\mathcal{Z}}(t)$.
This new probability is defined  as 
\begin{equation}
	p(\tau, j | \vec{\mathfrak{Z}}, t) \equiv a_j(\vec{\mathfrak{Z}}) \exp(-a_0(\vec{\mathfrak{Z}})\tau),
\end{equation}
where
\begin{equation}
	a_0(\vec{\mathfrak{Z}}) \equiv \sum_{j' = 1}^M a_{j'}(\vec{\mathfrak{Z}}).
\end{equation}
The probability is a joint probability density function of the two random variables "time to the next reaction", $\tau$:
\begin{equation}
	p(\tau | \vec{\mathfrak{Z}}, t) = a_0(\vec{\mathfrak{Z}}) \exp(-a_0(\vec{\mathfrak{Z}}) \tau),
\end{equation}
and "index of next reaction", $j$:
\begin{equation}
	p(j | \vec{\mathfrak{Z}}, t) = \frac{a_j(\vec{\mathfrak{Z}})}{a_0(\vec{\mathfrak{Z}})}.
\end{equation}
This probability can now be evaluated for every particle in a cell that fulfills the energetic requirement of a given reaction.

\subsection{Conservation of Moments}
%For a model to be physically applicable it needs to conserve relevant moments of the distribution function.
%In this work we focus only on the lower moments, namely \emph{mass}, \emph{momentum}, and \emph{energy}.
%To be fully consistent with the cubic FP approach one needs to conserve moments up to the heat flux.
%However, in this paper we only focus on the first 3 moments.
%The derivation of a method to conserve higher moments is left for future work.

For a model to be physically applicable, it must conserve the relevant moments of the velocity distribution function. In this work, we focus exclusively on the lower moments: \emph{mass}, \emph{momentum}, and \emph{energy}.
To achieve full consistency with the cubic FP approach, conservation of moments up to the heat flux would be required. However, in this paper, we limit our attention to the first three moments.
As a result, regions of the flow where a significant portion of particles engage in chemical reactions could lead to inaccuracies in the heat flux calculation.
However, for most types of flows in space engineering applications, such as those around space entry vehicles at very high Mach numbers and exhaust flows from space propulsion devices outside the combustion chamber, the proportion of reacted particles to total particles in a simulation cell is anticipated to be small.
Consequently, the development of a method to conserve higher-order moments will be addressed in future work.

\subsubsection{Mass conservation}
To ensure a strict mass conservation a mass balance is kept for each species $s_i$ present in a cell as outlines in Tab.~\ref{tab:mass_balance}.
\begin{table}[h!]
\caption{\label{tab:mass_balance}Mass Balance}
\centering
 \begin{tabular}{c | c | c} 
 \hline \hline
 Species & Available particles & Reacted particles \\ 
 \hline
 $s_1$  & $N_{\mm{av}}(s_1)$ & $N_{\mm{r}}(s_1)$  \\ 
 $s_2$ & $N_{\mm{av}}(s_2)$ & $N_{\mm{r}}(s_2)$  \\
 \vdots & \vdots & \vdots  \\
 $s_N$ & $N_{\mm{av}}(s_N)$ & $N_{\mm{r}}(s_N)$ \\ 
 \hline \hline
 \end{tabular}
\end{table}
This balance tracks the number of available particles for each species $N_{\mm{av}}(s_i)$, as well as the number of particles of each species that have been chosen for reaction but not yet accounted for, $N_{\mm{r}}(s_i)$.
At the beginning of each iteration, the number of particles for each species is calculated.
As the loop processes the particles in the cell, the count of available particles for a given species is reduced by one whenever a particle of that species is encountered.
Each time a reaction occurs, the counter $N_{\mm{r}}$  for the reacted partner species $s_p$ is incremented by one.
When a particle of species $s_p$ is next encountered, it is removed from the particle list, and the counter is decremented accordingly.

This approach ensures that reactions occur only when sufficient particles are available.
Particles that have already participated in a reaction are removed and prevented from reacting multiple times within the same time step.

\subsubsection{Momentum conservation}
The conservation of momentum is done on an integral basis.
This means the momentum is conserved on average per particle and is strictly conserved per cell.
The condition for this conservation is that the sum of the momenta of all reacted particles, $\vec{p}_e$  is equal to the sum of the momentum of the products of each reaction, $\vec{p}_p$.
This requirement can be formulated as:
\begin{equation}
    \label{eqn:moment_consrv}
    \sum_e \vec{p}_e \overset{!}{=} \sum_p \vec{p}_p.
\end{equation}
As we already have ensured mass conservation, Eqn.~(\ref{eqn:moment_consrv}) is equivalent to the conservation of the centre of mass velocity $\vec{v}_{\mm{cm}}$ between the reactants and the products of each reaction that took place:
\begin{equation}
    \label{eqn:vcm_consrv}
    \vec{v}_{\mm{cm}, e} \overset{!}{=} \vec{v}_{\mm{cm}, p}.
\end{equation}
To satisfy Eqn.~(\ref{eqn:vcm_consrv}) each product is assigned $\vec{v}_{\mm{cm}, e}$ as their base velocity. 

\subsubsection{Energy conservation}
To perform endothermic reactions, only particles with sufficient energy should be selected to participate.
Since the energies of all reaction partners are not known during the selection process, particles are chosen based on the information available for the selected particle in the loop and the average energy values calculated for each species.

For a given particle, the energy available for a reaction is assumed to be the sum of its internal and thermal energy, along with the average thermal and internal energy of the background
\begin{equation}
	e_{\mm{react}} = e_{\mm{therm}, p} + e_{\mm{int}, p} + \overline{e}_{\mm{kin}, b} + \overline{e}_{\mm{int}, b}.
\end{equation}
Here the subscript $p$ denotes the quantities associated with the particles, whereas subscript $b$ denotes quantities associated
with the background.
The individual terms can be expressed as:
\begin{equation}
	e_{\mm{react}} = \frac{1}{2} m_p c_p^2 + e_{\mm{int}, p} + \frac{3}{2} k_B T_b + \overline{e}_{\mm{int}, b}.
\end{equation}
where $c_p = v_p - \overline{v}_{\alpha}$ is the thermal velocity of a particle $p$ of species $\alpha$.
The total energy balance can then be written as:
\begin{equation}
	E_R + E_{\mm{therm}, c} + E_{\mm{int}, c} \overset{!}{=} E_P + E'_{\mm{therm}, c} + E'_{\mm{int}, c} + \Delta E.
\end{equation}
Here, $E_R$  is the total energy of all particles selected for chemical reactions. $E_P$ represents the total energy of the reaction products, including particles that participated in the reaction but did not themselves react.
$\Delta E$ is the total energy released or absorbed by all reactions during the time step.
$E_{\mm{therm}, c}$ and $E_{\mm{int}, c}$ denote the total thermal and internal energies of all other particles in the simulation cell.

Due to the momentum conservation constraints, the center of mass velocity (and consequently the energy) of the reaction products is determined by the center of mass velocity of the particles selected for the reaction. 
The energy balance can therefore be reformulated as:
\begin{equation}
	E_R + E_{\mm{therm}, c} + E_{\mm{int}, c} \overset{!}{=} E'_{\mm{coll}} + E'_{\mm{cm}} + E'_{\mm{therm}, c} + E'_{\mm{int}, c} + \Delta E
\end{equation}
where $E'_{\mm{cm}} = \frac{1}{2} v_{\mm{cm}}^2 \sum_r m_r$ is the total center of mass energy of all reaction products, and $E_{\mm{coll}}$ denotes their remaining collision energy.

The total energy contribution of the chemical reactions to the energy balance $\Delta E$ is determined by the number and type of reactions that have taken place. However, since particles are selected partly based on average energy assumptions, a correction for these assumptions is required.
To maintain the energy balance, we adjust the thermal energy of the particles in the cell after the reactions have occurred, while the internal energies of non-reacting particles remain unchanged.
This is achieved by introducing a linear scaling factor $q$ for the energy, given by:
\begin{equation}
	E_R + E_{\mm{therm}, c} \overset{!}{=} \underbrace{E'_{\mm{coll}}}_{q (E_R - E'_{\mm{cm}})} + E'_{\mm{cm}} + \underbrace{E'_{\mm{therm}, c}}_{q E_{\mm{therm}, c}} + \Delta E.
\end{equation}
After all reactants have been selected, the scaling factor $q$ can be calculated as:
\begin{equation}
q = 1 - \frac{\Delta E}{E_R - E'_{\mm{cm}} + E_{\mm{therm}}}
\end{equation}
where $E_R - E'_{\mm{cm}}$ represents the actual total available reaction energy of the selected particles.

\subsection{Energy of created particles}
The chemical process can produce both monoatomic and polyatomic species.
Therefore, the correct energy distribution among the products must be modeled. 
To achieve this, the remaining collision energy, $E_{\mm{coll}}$ is distributed as follows:
\begin{enumerate}
	\item The products' internal energy is sampled from the Boltzmann distribution with temperature $T^{\mm{eq}}$ which is assumed to be the same for all products.
	\item Collision energy $\tilde{E}_{\mm{coll}}$ is updated as $\tilde{E}_{\mm{coll}} = E_{\mm{coll}} - \sum_p e_{p, \mm{int}}(T^{\mm{eq}})$
	\item The energy $\tilde{E}_{\mm{coll}}$ is used to calculate the ``thermal velocity'' of the products in the same manner as in DSMC for elastic collisions.
\end{enumerate}

To calculate the equilibrium temperature  $T^{\mm{eq}}$, we assume that the collision energy can be expressed as:
\begin{equation}
	E_{\mm{coll}} = e_{\mm{therm}}(T^{\mm{eq}}) + \sum_p e_{p, \mm{rot}}(T^{\mm{eq}}) + \sum_p e_{v, \mm{rot}}(T^{\mm{eq}}).
\end{equation}

\subsubsection{Thermal Energy}
The thermal energy is taken to be
\begin{equation}
e_{\mm{therm}}(T^{\mm{eq}}) = \frac{1}{2} m_r \bigg( \langle |\vec{g}| \rangle(T^{\mm{eq}}) \bigg)^2,
\end{equation}
where $\langle |\vec{g}|^2 \rangle(T^{\mm{eq}})$ is the average relative velocity as a function of the equilibrium temperature $T^{\mm{eq}}$ and $m_r$ is the reduced mass of the reacting species.
The average relative velocity can be calculated as
\begin{equation} 
	\langle |\vec{g}| \rangle = \int \int |\vec{v}_1 - \vec{v}_2| f_1(\vec{v}_1) f_2(\vec{v}_2) \diff{\vec{v}_1} \diff{\vec{v}_2} = \sqrt{\frac{8 T \kb}{\pi m_r}}
\end{equation}
which leads to the expression for the thermal energy in the form:
\begin{equation}
	e_{\mm{therm}}(T^{\mm{eq}}) = \frac{4}{\pi} \kb T^{\mm{eq}}.
\end{equation}

\subsubsection{Rotational Energy}
To calculate the equilibrium temperature of the products, fully excited rotational degrees of freedom are assumed. Therefore the rotational energy can be written as:
\begin{equation}
	e_{\mm{rot}}(T^{\mm{eq}}) = \frac{\mm{dof}_{\mm{rot}}}{2} \kb T^{\mm{eq}},
\end{equation}
with $\mm{dof}_{\mm{rot}}$ denoting the number of rotational degrees of freedom.

\subsubsection{Vibrational Energy}
The vibrational energy of the particles is modeled with an infinite harmonic oscillator:
\begin{equation}
	\label{eqn:evib}
	e_{\mm{vib}}(T^{\mm{eq}}) = \kb T^{\mm{eq}} \sum_i g_i \frac{\theta_i/ T^{\mm{eq}}}{\exp(\theta_i/T^{\mm{eq}}) - 1},
\end{equation}
where $\theta_i$ is the characteristic vibrational temperature and $g_i$ is the degeneracy of the $i$-th vibrational mode.
Since Eq.~(\ref{eqn:evib}) cannot be solved directly for $T^{\mm{eq}}$ a linearization is used instead in the form:
\begin{align}
	e_{\mm{vib}}(T^{\mm{eq}}; T_0) &=  e_{\mm{vib}}(T_0) +  \pdv{e_{\mm{vib}}}{T^{\mm{eq}}} \bigg|_{T_0} (T^{\mm{eq}} - T_0)\\
&= \underbrace{\pdv{e_{\mm{vib}}}{T^{\mm{eq}}} \bigg|_{T_0}}_{\partial e_{\mm{vib}}} T^{\mm{eq}} + \underbrace{e_{\mm{vib}}(T_0) - \pdv{e_{\mm{vib}}}{T^{\mm{eq}}} \bigg|_{T_0} T_0}_{q_{\mm{vib}}}\\
&= \partial e_{\mm{vib}} \times T^{\mm{eq}} + q_{\mm{vib}}.
\end{align}
For the linearization point the classical assumption is made that energy is distributed with $\frac{1}{2}\kb T$ per degree of freedom which leads to the equation
\begin{equation}
	E_{\mm{coll}} =  \kb T \left[ \frac{4}{\pi} + \frac{1}{2}\sum_p \left(\mm{dof}_{\mm{rot}, p} + \mm{dof}_{\mm{vib}, p} \right) \right],
\end{equation}
which, using $\mm{Dof} = \sum_p \mm{dof}_{p}$, can be solved for $T_0$ as
\begin{equation}
	T_0 = \frac{E_{\mm{coll}}}{\kb \left[ \frac{4}{\pi} + \frac{1}{2}  \mm{Dof}_{\mm{rot}} + \frac{1}{2}  \mm{Dof}_{\mm{vib}}  \right]}.
\end{equation}

\subsubsection{Total Energy}
Using the definition of the linearized vibrational energy $e_{\mm{vib}}(T; T_0)$ the collision energy can be written as
\begin{align}
	E_{\mm{coll}}  &= \frac{4}{\pi} \kb T + \sum_p \left[\frac{\mm{dof}_{\mm{rot}, p}}{2} \kb T + \partial_T e_{\mm{vib}, p} \times T + q_{\mm{vib}, p} \right] \\
	&= \frac{4}{\pi} \kb T + \frac{1}{2}\mm{Dof}_{\mm{rot}} \kb T + \partial_T E_{\mm{vib}} \times T + Q_{\mm{vib}}.
\end{align}
where $\partial_T E_{\mm{vib}} = \sum_p \partial_T e_{\mm{vib}, p}$ and $Q_{\mm{vib}} = \sum_p q_{\mm{vib}, p}$.
Using this derivation we can now calculate the equilibrium temperature 
\begin{equation}
	\label{eqn:T_eq}
	T^{\mm{eq}} = \frac{E_{\mm{coll}} - Q_{\mm{vib}}}{\frac{4}{\pi}\kb  + \frac{1}{2}\kb  \mm{Dof}_{\mm{rot}} + \partial_T E_{\mm{vib}}}.
\end{equation}
The remaining energy which is split between the created particles to calculate their kinetic energy can now be expressed as
\begin{equation}
	\label{eqn:E_coll}
	\tilde{E}_{\mm{coll}} = E_{\mm{coll}} - \sum_p \big\{ e_{\mm{rot}, p}(T^{\mm{eq}}) + e_{\mm{vib}, p}(T^{\mm{eq}}) \big\}.
\end{equation}

\subsection{Considered Reaction Types}
The proposed chemistry model is independent of the specific reaction types and rate coefficients used. In this paper, we focus on reactions that are relevant in the context of shocks in hypersonic flows, specifically \emph{dissociation} and \emph{exchange} reactions.

\subsubsection{Exchange Reactions}
Probably the most straightforward reaction type to model in particle based methods, the exchange reaction describes the exchange of a component \ch{B} of a polyatomic molecule \ch{AB} with a collision partner \ch{C} given a rate constant $k$ in the form
\begin{equation}
    \ch{AB + C ->[\textit{k}] A + CB}.
\end{equation}
Following the approach described by Bird \cite{bird_molecular_1994}, to conserve momentum the products are assigned the velocities
\begin{align}
	\vec{v}_{\ch{A}} &= \vec{v}_{\mm{cm}} + \frac{m_{\ch{CB}}}{m_{\ch{A}} + m_{\ch{CB}}} \vec{g},\\
	\vec{v}_{\ch{CB}} &= \vec{v}_{\mm{cm}} - \frac{m_{\ch{A}}}{m_{\ch{A}} + m_{\ch{CB}}} \vec{g}, 
\end{align}
where $\vec{g}$ is the relative (thermal) velocity vector which orientation is randomly distributed on a unit sphere.
The magnitude of $\vec{g}$ is computed as
\begin{equation}
	\left|\vec{g}\right| = \sqrt{\frac{\tilde{E}_{\mm{coll}}}{m_r}},
\end{equation}
with $\tilde{E}_{\mm{coll}}$ being the approximated collision energy defined in Eqn.~(\ref{eqn:E_coll}) and $m_r$ the reduced mass
\begin{equation}
	m_r = \frac{m_{\ch{A}} m_{\ch{CB}}}{m_{\ch{A}} + m_{\ch{CB}}}.
\end{equation}
Created particles belonging to polyatomic species are assigned rotational and vibrational energies based on on $T^{\mm{eq}}$ defined in Eqn.~(\ref{eqn:T_eq}).
%\hl{Here I see the issue that "macroscopic velocities would not be conserved"}.
% \hl{Maybe I will add a reference to my chemistry paper with Manuel --- we discussed different velocity choices there... - G.}
For a discussion of other possible post-collisional velocity choices, the reader is referred to~\cite{sarna2021moment}.
%An example is a charged beam that undergoes charge exchange via collisions with "background" neutral gas.
%In this way the charge is transmitted to slow particles that kept trapped i.e. in a charge exchange thruster and the now neutral fast particles leave the system.

\subsubsection{Dissociation Reactions}
Dissociation reactions describe the decomposition of a polyatomic molecules into its components.
In this paper we focus on collision induced dissociation with a polyatomic molecule \ch{AB} breaking into two components via a collision with an interaction partner \ch{M}:
\begin{equation}
    \ch{M + AB ->[\textit{k}] M + A + B}.
\end{equation}
The process is very similar to the one described for exchange reactions.
The difference is that now three particles are created instead of two.
In this case both the products \ch{A} and \ch{B} are assigned the same thermal post-reaction velocity calculated as
\begin{align}
	\vec{v}_{\ch{M}} &= \vec{v}_{\mm{cm}} + \frac{m_{\ch{AB}}}{m_{\ch{M}} + m_{\ch{AB}}} \vec{g}, \\
	\vec{v}_{\ch{A}} &= \vec{v}_{\mm{cm}} - \frac{m_{\ch{M}}}{m_{\ch{M}} + m_{\ch{AB}}} \vec{g}, \\
	\vec{v}_{\ch{B}} &= \vec{v}_{\ch{A}}.
\end{align}
Again the internal energies of the products can be sampled using $T^{\mm{eq}}$.
\section{Verification}
In this section we provide verification results for our chemistry model based on comparisons to analytical solutions as well as 0D  simulation results created using multi-temperature continuum models as well as the SPARTA DSMC code \cite{plimpton_direct_2019}.
For the verification process, unless otherwise specified, we use Arrhenius-type rate constants $k$ of the form:
\begin{equation}
	\label{eqn:arrhenius}
	k(T) = A T^B \exp\left(-\frac{E_a}{\kb T} \right)
\end{equation}
where $A$ and $B$ are the Arrhenius coefficients, $T$ is the kinetic temperature, and $E_a$ is the activation energy.
Our proposed chemistry model is agnostic to the choice of rate constant. This means that other formulations of rate constants, such as Bird's Quantum-Kinetic model (QK) \cite{bird_q-k_2011} or tabulated values, can be used. Since the FP model is valid near the continuum regime, we use the Arrhenius equation for consistent comparisons with established Navier-Stokes models.

\subsection{Calculation of Rate Constants}
\label{sec:calculation_of_rate_constants}
As the first step of the verification process, we demonstrate that our model correctly reproduces the provided rate constants. To achieve this, a 0D FP simulation is run multiple times for selected temperature points using different random seeds. The number of evaluated reactions is then averaged over all executions and used to calculate the "effective" rate constant produced by our simulation. This rate constant is subsequently compared to the rate constant predicted by the Arrhenius model, as shown in Eqn.~(\ref{eqn:arrhenius}).

The chemical process in this case is a system of two coupled dissociation reactions
\begin{align}
    \ch{O2 + O2}&\overset{\textit{r} 1}{\longrightarrow}\ch{O2} + 2\ch{O}\\
    \ch{O + O2}&\overset{\textit{r} 2}{\longrightarrow}3\ch{O}
% 	\ch{O2 + O2 &->[\textit{r} 1] O2 + 2 O}\\
% 	\ch{O + O2 &->[\textit{r} 2] 3 O}
\end{align}
with the reaction rates being
\begin{align}
	r_1 &= k_1(T)[\ch{O_2}][\ch{O_2}]\\
	r_2 &= k_2(T)[\ch{O}][\ch{O_2}].
\end{align}
The simulation domain is a cube with a volume of $V = \SI{e-12}{\cubic\meter}$. 
We used a time step of $\Delta t = \SI{e-10}{\second}$, and a particle weight of $w=\num{e5}$.
The number densities at the beginning of the simulation where $n_{\ch{O}}(t_0) = \SI{e21}{\per\meter\cubed}$ and $n_{\ch{O2}}(t_0) = \SI{9.9e22}{\per\meter\cubed}$.
The used Arrhenius parameters can be found in Tab.~\ref{tab:air_chemistry_o2_diss}.
%\begin{table}[h!]
%\caption{\label{tab:reaction_rates_setup}Arrhenius Parameters for Coupled \ch{O2 + O} Dissociation}
%\centering
 %\begin{tabular}{l|ccc} 
%Reaction &  $A$ / $\left(\si{\meter\cubed\per\second\kelvin\tothe{-B}}\right)$  & $B$ & $E_a$ / $\si{\joule}$\\
 %\hline
%\ch{O2 + O2 -> O2 + 2 O} & \num{3.321e-9} & \num{-1.5} & \num{8.197e-19}\\
%\ch{O + O2 -> 3 O} & \num{1.660e-8} & \num{-1.5} & \num{8.197e-19}\\
% \end{tabular}
%\end{table}
The results for each temperature point were averaged over 100 samples.

\begin{figure}
\centering
\begin{subfigure}{0.45\textwidth}
\centering
\includegraphics[width = \textwidth]{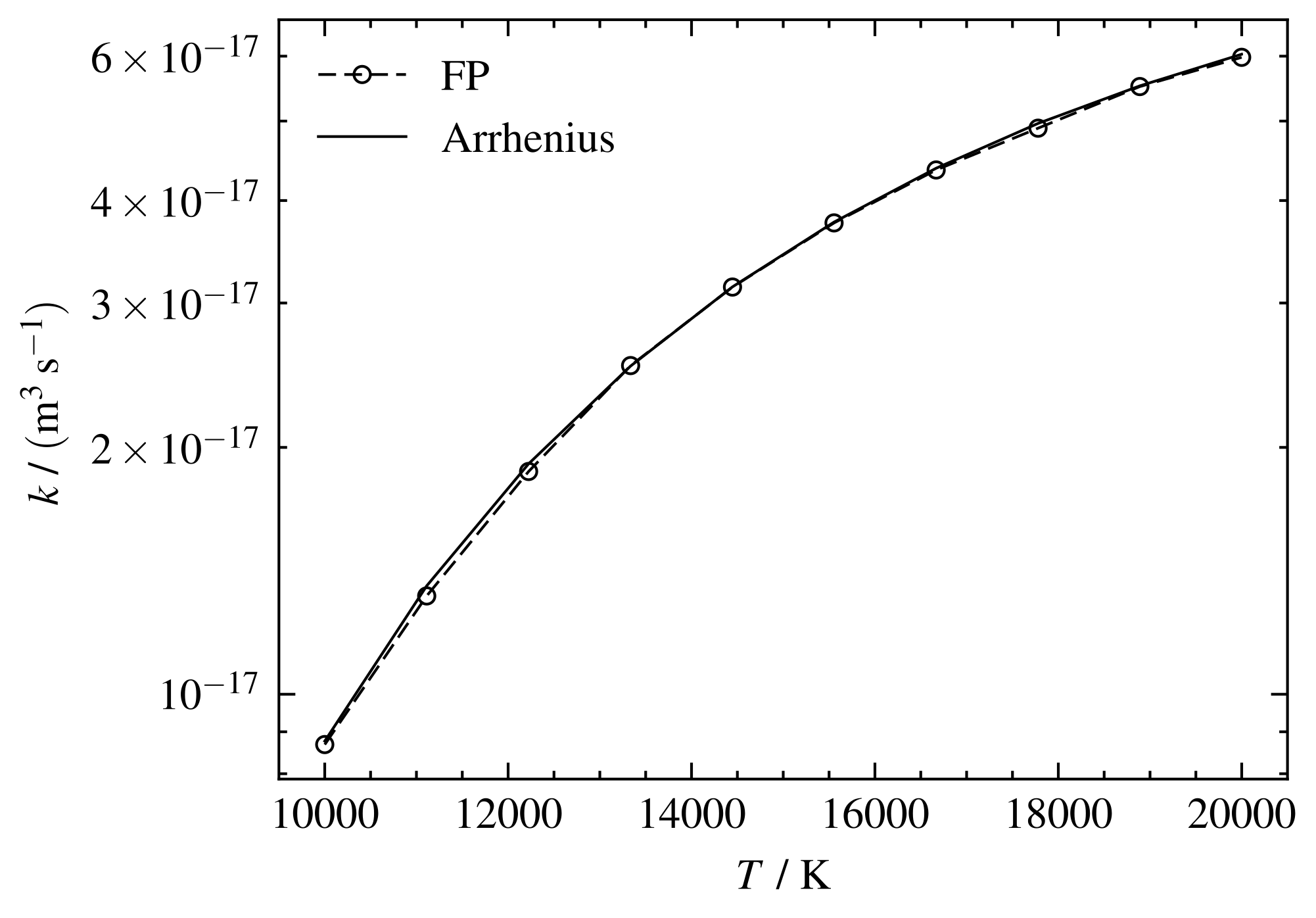}
\caption{\ch{O2 + O2 -> O2 + 2 O}}
\label{fig:left}
\end{subfigure}\vspace{0.03\textwidth}
\begin{subfigure}{0.45\textwidth}
\centering
\includegraphics[width = \textwidth]{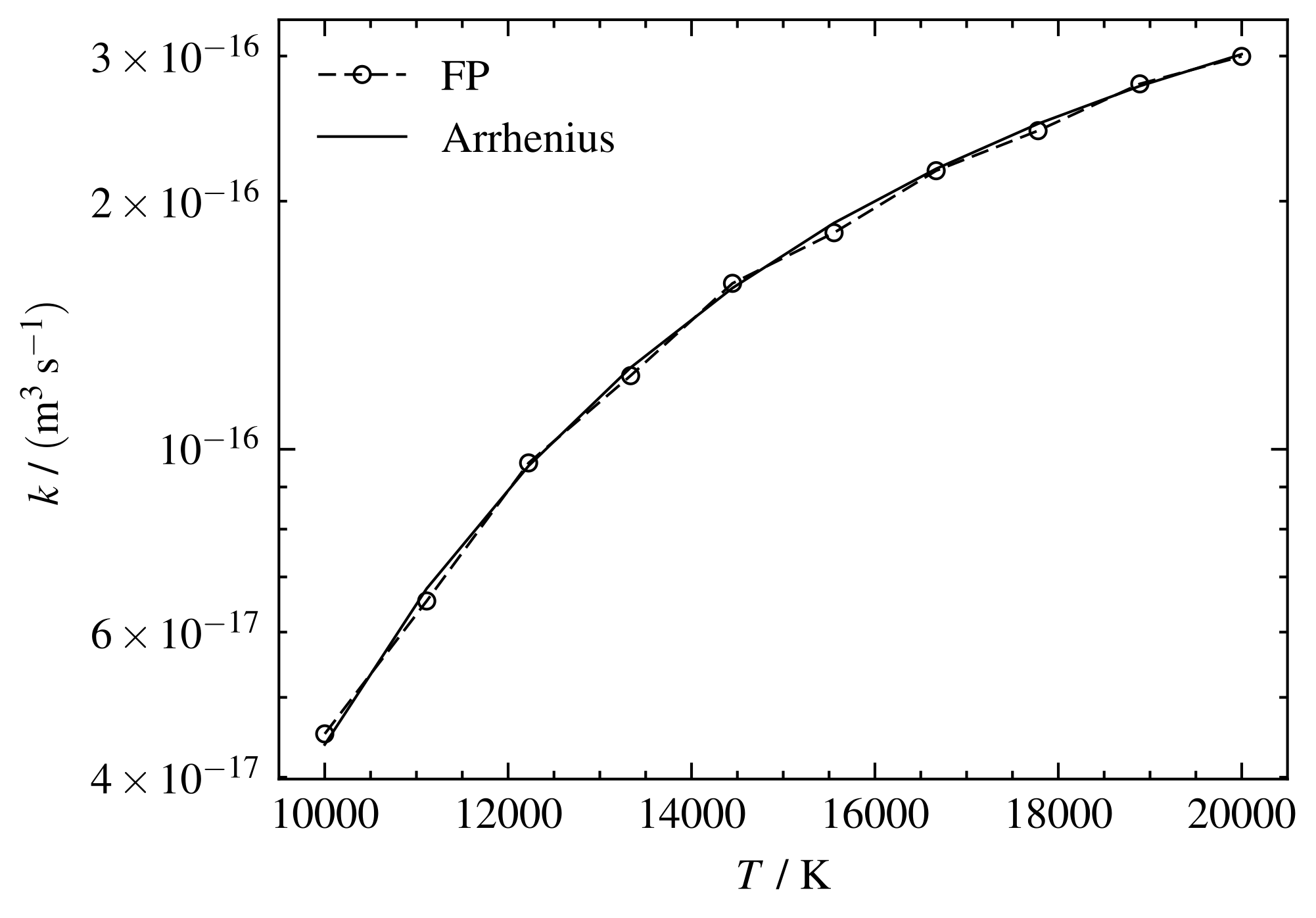}
\caption{\ch{O + O2 -> 3 O}}
\label{fig:right}
\end{subfigure}
\caption{Combined dissociation of \ch{O2}.}
\label{fig:rate_const_combined_dissocitaion}
\end{figure}

%For exchange reactions we have Fig.~(\ref{fig:rate_const_NO_O}).
%\begin{figure}
%    \includegraphics[width=0.6\textwidth]{figures/NO_O_compare_FP_Arr_dt_1e-07_fnum_10000000.0_nrho1e+23.png}
%    \caption{\label{fig:rate_const_NO_O} Rate constant for $\ch{N2 + N}$ comparison calculated using the Arrhenius equation and reproduced from FP simulations.}
%\end{figure}
The evaluated results of the simulations can be seen in Fig.~(\ref{fig:rate_const_combined_dissocitaion}).
The rate constants calculated from the FP simulations match very well the ones predicted by the Arrhenius equation showing that the evaluation of the reaction probabilities is correctly implemented.

\subsection{Reaction Kinetics, $k = \mm{const}$}
To investigate whether the chemical kinetics are reproduced correctly, as well as if the lower moments are conserved as expected,  we consider a dissociation reaction of diatomic nitrogen 
\begin{equation}
\ch{N2 + N ->[\textit{k}] 3 N},
\end{equation}
where the reaction rate has the form of
\begin{equation}
    r = k [\ch{N2}] [\ch{N}].
\end{equation}
Assuming an isolated system with a constant volume, a temperature-independent rate constant, and the reaction being neither endo- nor exothermic, an analytic solution for the time dependent number density can be found 
\begin{equation}
    n_{\ch{N2}}(t) = \frac{n_{\ch{N2}}(t_0) Q}{2 n_{\ch{N2}}(t_0) - \big(n_{\ch{N2}}(t_0) - n_{\mm{tot}}(t_0) \big) \exp\big(Q\,k\,t\big)}
\end{equation}
where
\begin{equation}
    Q = \big(n_{\ch{N2}}(t_0) + n_{\mm{tot}}(t_0) \big)
\end{equation}
and
\begin{equation}
    n_{\mm{tot}}(t_0) = n_{\ch{N2}}(t_0) + n_{\ch{N}}(t_0).
\end{equation}
For the verification test case a rate constant of $k=\SI{e-15}{\per\second\meter\tothe{6}}$ was chosen.
The remaining simulation parameters are summarized in Tab.~\ref{tab:values_analytical_dissociation}.
\begin{table}[h!]
\caption{\label{tab:values_analytical_dissociation}Simulation Parameters Used for Analytical Dissociation Test Case}
\centering
 \begin{tabular}{ccccc} 
$\Delta t / \si{\second}$ & $V / \si{\meter\cubed}$ & $w$ & $n_{\ch{N}}(t_0) / \si{\per\meter\cubed}$ & $n_{\ch{N2}}(t_0) / \si{\per\meter\cubed}$\\
 \hline
    \num{e-10} & \num{e-12} & \num{e6} & \num{2e20} & \num{9.8e21}\\
 \end{tabular}
\end{table}

\begin{figure}
    \includegraphics[width=0.6\textwidth]{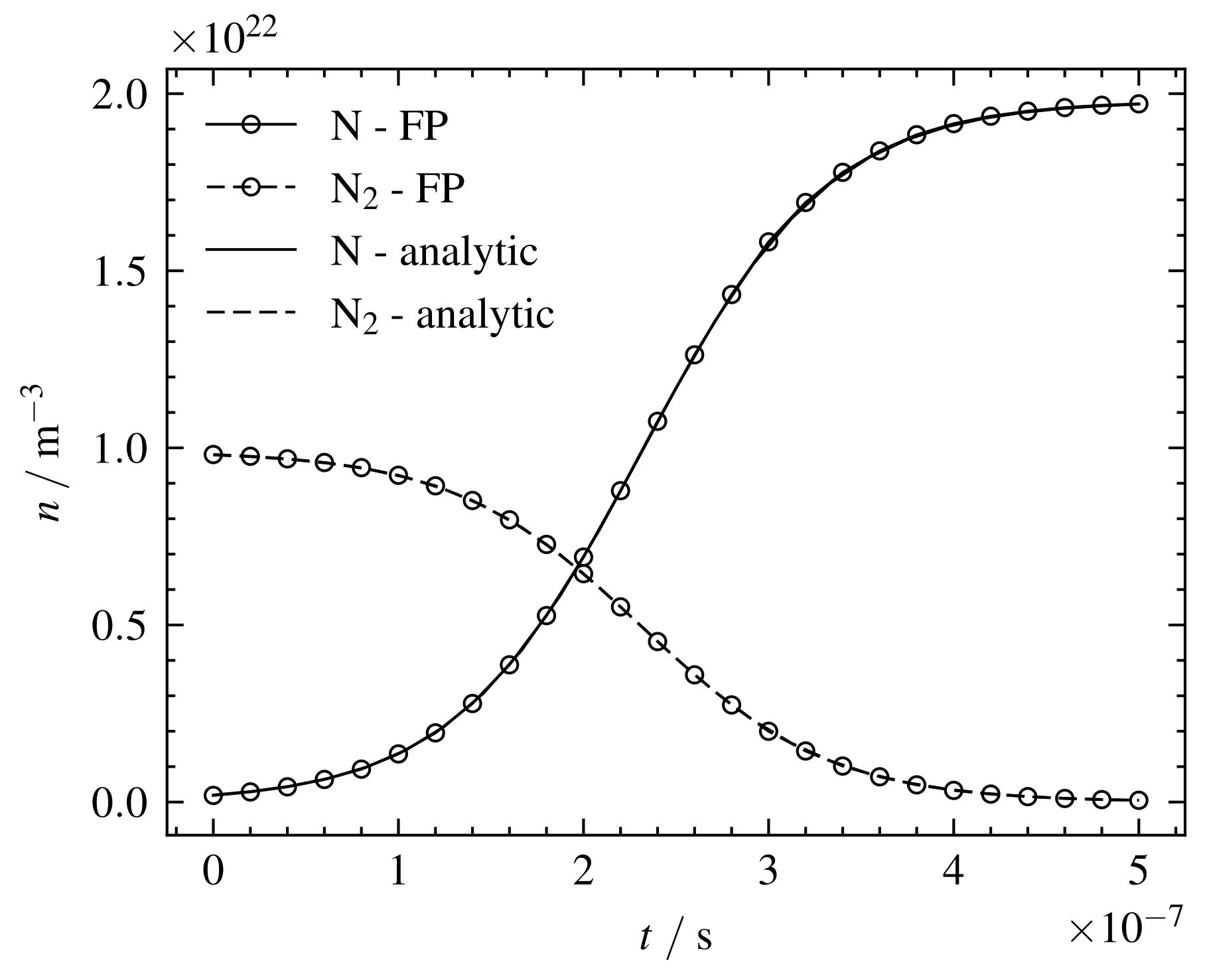}
    \caption{\label{fig:analytic_dissociation} Chemical kinetics for \ch{N2 + N} dissociation using a constant rate constant of $k=\SI{1e-15}{\per\second\meter\tothe{6}}$.}
\end{figure}

\begin{figure}
    \includegraphics[width=0.6\textwidth]{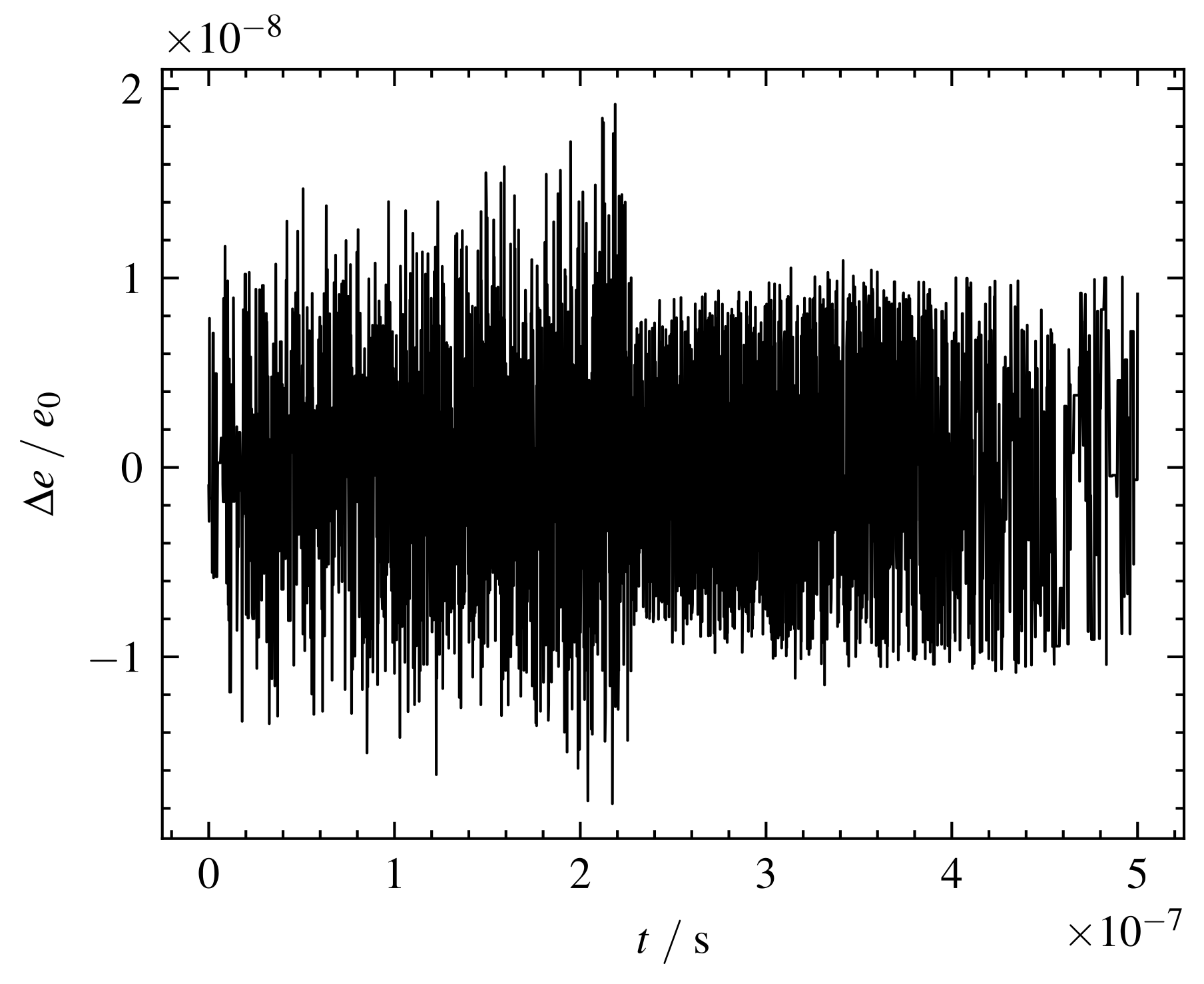}
    \caption{\label{fig:analytic_dissociation_energy_conserv} Energy conservation.}
\end{figure}

The simulation was performed 5 times, each time using a different seed for random numbers.
The results  displayed in Fig.~(\ref{fig:analytic_dissociation}) are averaged over these 5 simulations.
Very good agreement is observed between the analytic result and the averaged simulation results produced by our chemistry model.
Mass and momentum are exactly conserved. %as can be seen in Fig.~(\ref{fig:analytic_dissociation_mass_conserv})  and Fig.~(\ref{fig:analytic_dissociation_momentum_conserv}) respectively.
%\begin{figure}
%    \includegraphics[width=0.6\textwidth]{figures/Analytic_dissociation_mass.png}
%    \caption{\label{fig:analytic_dissociation_mass_conserv} Mass conservation.}
%\end{figure}
%\begin{figure}
%    \includegraphics[width=0.6\textwidth]{figures/Analytic_dissociation_momentum.png}
%    \caption{\label{fig:analytic_dissociation_momentum_conserv} Momentum conservation.}
%\end{figure}
The energy is conserved but the output is subject to noise as seen in Fig.~(\ref{fig:analytic_dissociation_energy_conserv}).
The reason for the noise is,  that energy is a second order moment of the velocity distribution function, coupled with SPARTA's output format for floating point values being restricted to 7 decimal digits by default.

\subsection{Reaction Kinetics, $k = f(T)$}
\label{sec:reaction_kinetcs_k_T}
Even for a spatially homogenous gas, in the presence of vibrational relaxation and/or multiple reactions, it is no longer possible to obtain an analytical solution for the system of master equations describing the thermochemical relaxation of the gas mixture. Therefore, a numerical solution of a multi-temperature system of master equations~\cite{nagnibeda2009non} as well as simulation results produced by the SPARTA DSMC code with the Total Collision Energy model (TCE) are used as reference solutions against which the proposed Fokker-Planck chemistry algorithm is verified.
We use the extension of the DSMC algorithm  introduced by Zhang and Schwartzentruber \cite{zhang_inelastic_2013} to model energy relaxation for gas mixtures.

The relaxation of internal degrees of freedom in the FP model was modeled by the master equation ansatz for VHS gas mixtures \cite{hepp_kinetic_2022}.
Temperature dependent relaxation collision numbers $Z$ for both rotational and vibrational relaxation where used \cite{bird_molecular_1994}.
Collision numbers for the relaxation  between two species $\alpha$ and $\beta$ are the averages between the individual species collision numbers
\begin{equation}
	Z_{\alpha, \beta}(T) = \frac{Z_{\alpha}(T) + Z_{\beta}(T)}{2}.
\end{equation}

We investigate exchange and dissociation reaction sets individually to verify the functionality.
For all cases the simulation domain for the DSMC and FP simulations is a cube with a side length of \SI{e-5}{\meter} and periodic boundary conditions.
The starting temperature of each simulation is chosen to be $T = \SI{10000}{\kelvin}$ with a total number density of $n = \SI{e23}{\per\meter\cubed}$.
The mole fraction of each species is set to be equal to $x = \num{0,2}$.
A particle weight of $w = \num{e-6}$ and a time step of $\Delta t = \SI{e-8}{\second}$ are used.

\subsubsection{Exchange}
We consider the exchange reactions presented in table~\ref{tab:air_chemistry_exchange} of the appendix.
%It must be noted that the selection energy for the reaction $\ch{O2 + N -> NO + O}$ is not the same as $E_a = 0$ but rather $\Delta E = \SI{2.2129e-19}{\joule}$.

\begin{figure}
    \includegraphics[width=0.6\textwidth]{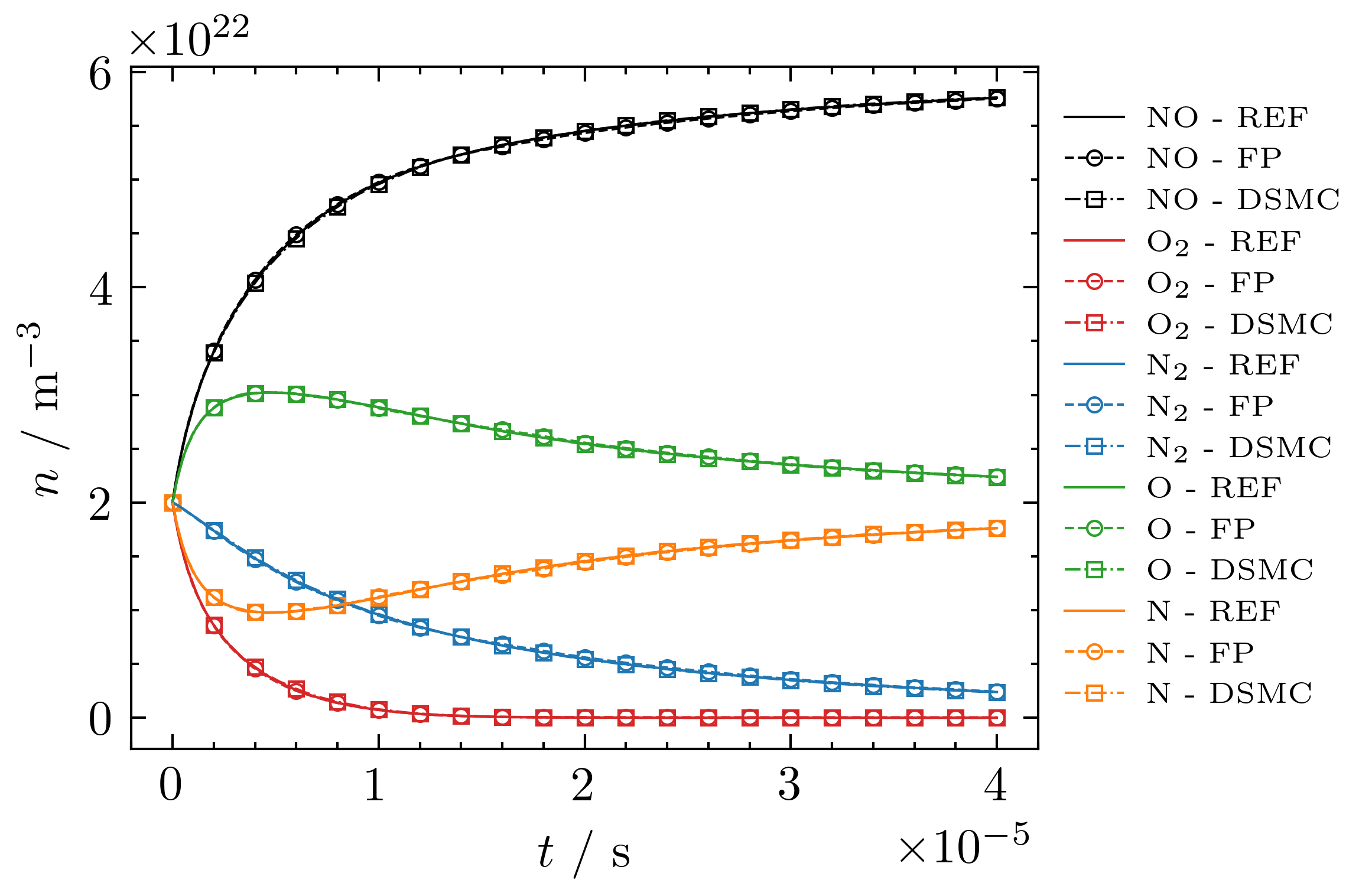}
    \caption{\label{fig:nrho_exchange_2_reactions} Number density using both exchange reactions.}
\end{figure}

\begin{figure}
    \includegraphics[width=0.6\textwidth]{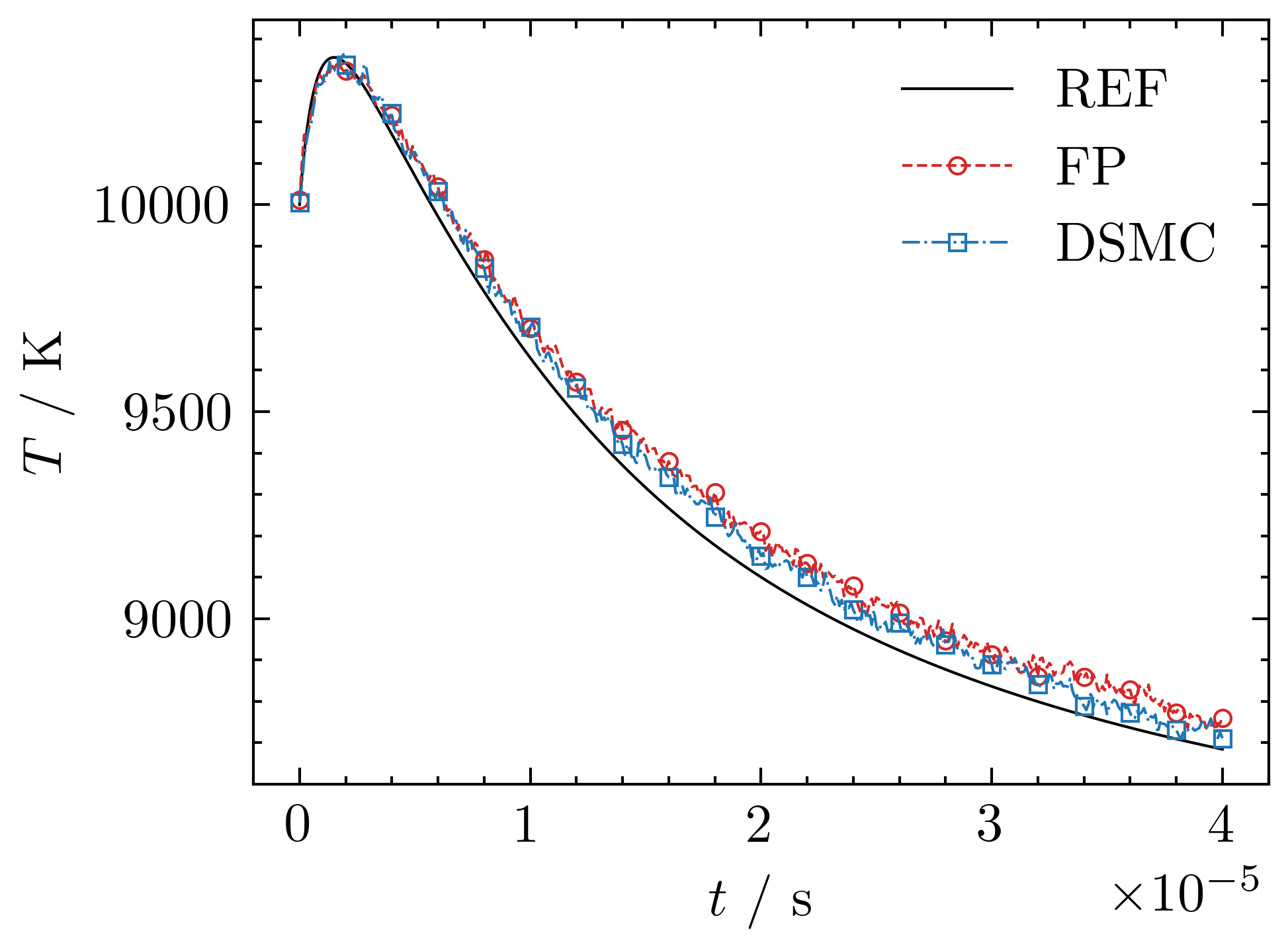}
    \caption{\label{fig:temp_exchange_2_reactions} Temperature using both exchange reactions.}
\end{figure}

Very good agreement between all models can be observed for the  temporal development of the number densities of the participating species as can be seen in Fig.~\ref{fig:nrho_exchange_2_reactions}.
In all figures, REF refers to the continuum, multi-temperature model.
While the temperature profile agrees very well between DSMC and FP, small deviations can be observed in relation to the multi-temperature model as seen in Fig.~\ref{fig:temp_exchange_2_reactions}.
These minor differences may be attributed to differences in the modelling of vibrational relaxation between DSMC and the continuum approaches.

\subsubsection{Dissociation}
\begin{figure}
    \includegraphics[width=0.6\textwidth]{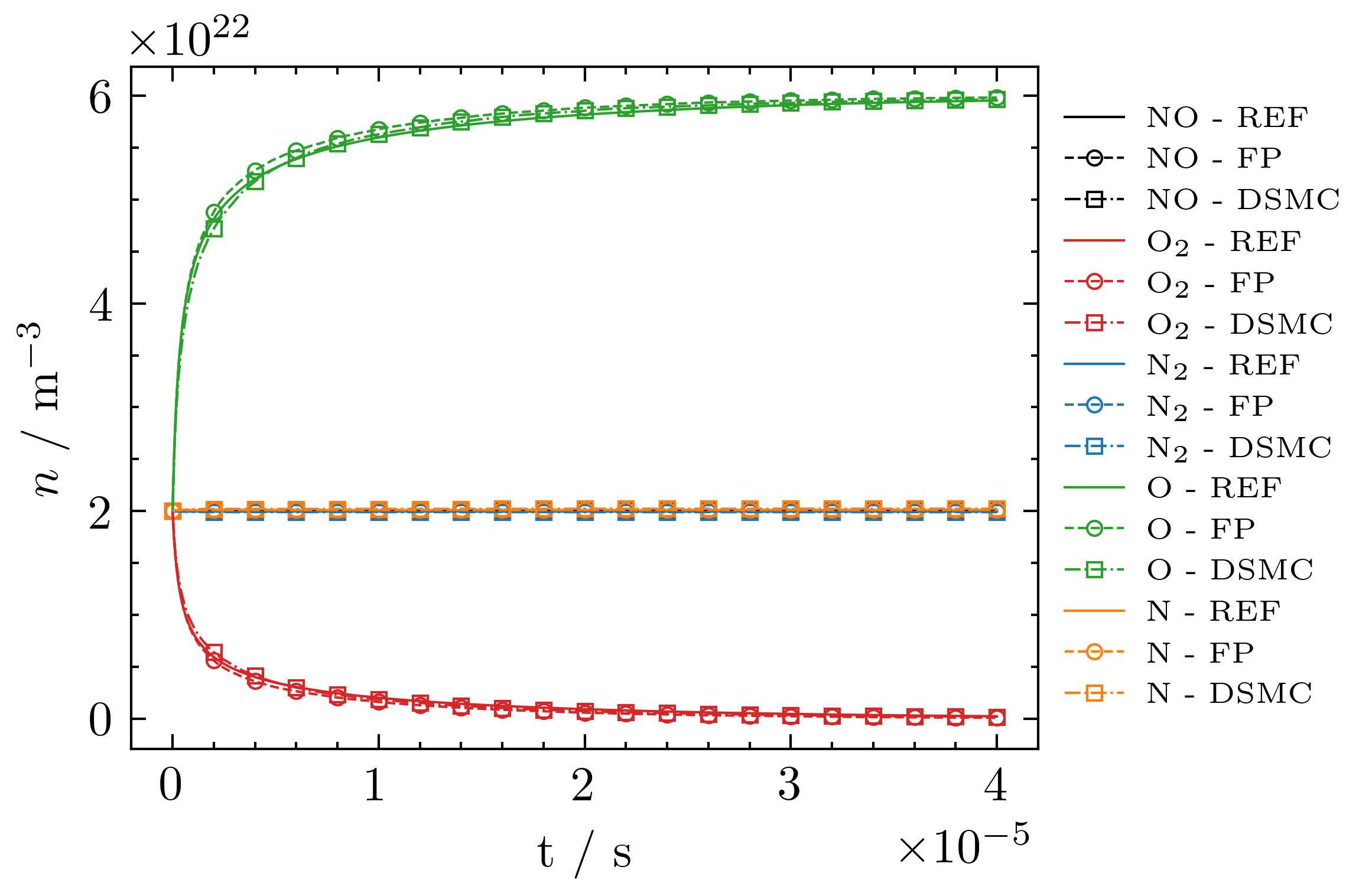}
    \caption{\label{fig:nrho_dissociation_10_reactions} Number density using 10 dissociation reactions.}
\end{figure}

\begin{figure}
    \includegraphics[width=0.6\textwidth]{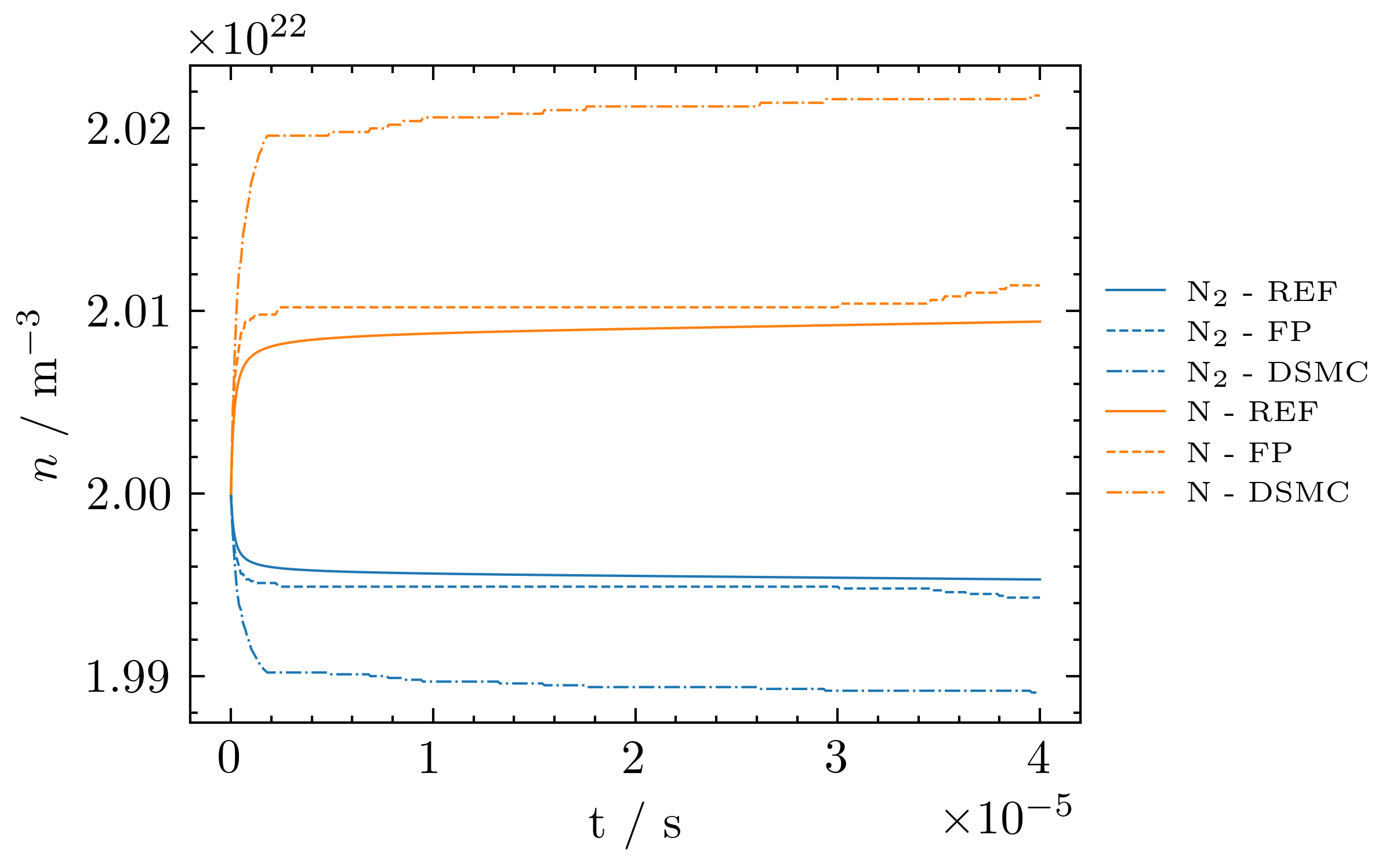}
    \caption{\label{fig:nrho_dissociation_10_reactions_N2_N} Number density of \ch{N2} and \ch{N} using 10 dissociation reactions (detail).}
\end{figure}

\begin{figure}
    \includegraphics[width=0.6\textwidth]{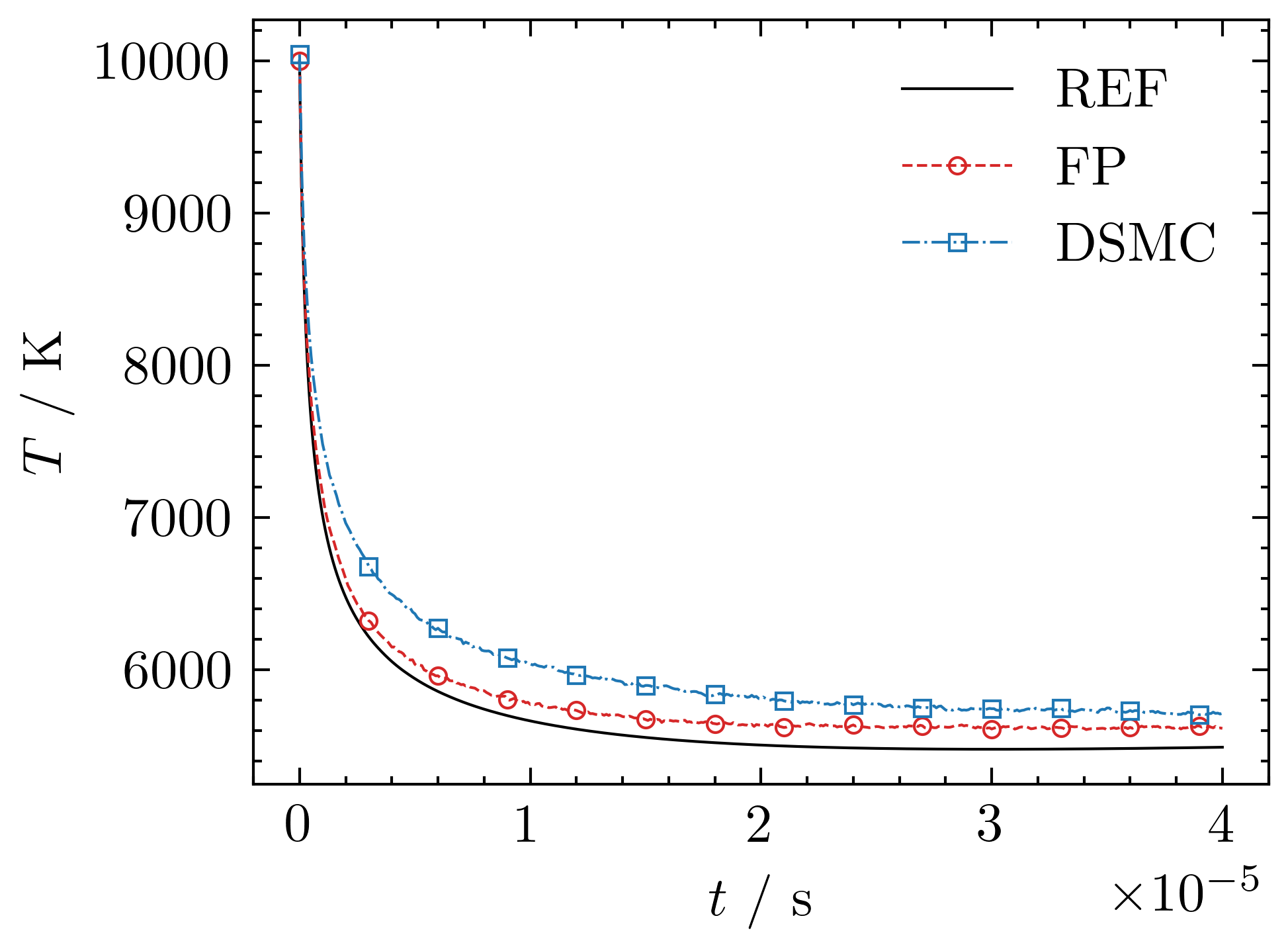}
    \caption{\label{fig:temp_dissociation_10_reactions} Temperature using 10 dissociation reactions.}
\end{figure}

For the dissociation test case \ch{O2} and \ch{N2} reactions are used from the appendix Tables~\ref{tab:air_chemistry_o2_diss} and \ref{tab:air_chemistry_n2_diss} respectively.
Good agreement can be observed between all models for \ch{O2} and \ch{O} kinetics (seen in Fig~\ref{fig:nrho_dissociation_10_reactions}).
The agreement is also very good between FP and the multi-temperature model as seen in Fig.~\ref{fig:nrho_dissociation_10_reactions_N2_N} and shows slightly larger deviations in respect to the DSMC results.
This shouldn't distract from the fact that the relative deviations are still small.
The same holds true for the temporal development of the translational temperature as displayed in Fig.~\ref{fig:temp_dissociation_10_reactions}.
The slightly larger differences between the models in the temperature profile compared to results from exchange reactions are expected as now a considerably larger number of reactions is considered with a much larger temperature drop over the run of the simulation.
Due to this rapid evolution of the system, the differences in the modeling of vibrational relaxation between the considered approaches are exacerbated, leading to more noticeable differences in the temperature.
In addition, the use of the temperature-dependent collision numbers with the Larsen--Borgnakke model may lead to the model not satisfying detailed balance \cite{eckert_enforcing_2022}, an issue which is not present in the continuum-based approach, thus further affecting the evolution of the temperature.
However, this shouldn't distract from the fact that the relative deviations both in the number densities and the temperature are still small. 
\section{Performance Analysis}
The advantage of the FP approach compared to DSMC is that its computational cost scales with the number of simulation particles rather than with the collision frequency, making the model computationally more efficient in regions with low Knudsen numbers. In this section, we investigate how the computational cost of our chemistry model scales as the Knudsen number decreases, and compare this scaling to the DSMC TCE model implemented in SPARTA.
%For the performance analysis the exchange reaction was performed with varying Kn numbers.
The simulation of the exchange reaction 
\begin{equation}
\ch{O2 + N -> O + NO}
\end{equation}
was performed in a fixed volume cube with the side length $L=\SI{e-4}{\meter}$ with specular reflective boundary conditions.
The initial temperature of the simulation was set to $T(t_0) = \SI{20000}{\kelvin}$.
The Knudsen number was modified by varying the initial number density of the mixture while retaining the mole fractions $x_{\ch{O2}} = x_{\ch{N}}(t_0) = \frac{1}{2}$.
The particle number was kept constant at $N = \num{100000}$ for all simulation by adjusting the particle weight.
The Knudsen number was computed as
\begin{equation}
	\mm{Kn} = \frac{\lambda_{\ch{O2}} + \lambda_{\ch{N}}}{2 L}
\end{equation}
with the mean free path $\lambda$ being calculated using the Variable Hard Sphere (VHS) model
\begin{equation}
	\lambda = \frac{1}{\sqrt{2}d_{\mm{ref}}^2 n \left( \frac{T_{\mm{ref}}}{T} \right)^{\omega - \frac{1}{2}}}.
\end{equation}
Each simulation was executed for  100 time steps.
The results were averaged over 100 simulations each and normalized with the average execution time of a FP simulation $\overline{t_{\mm{FP}}} = \sum_i^N t_{\mm{FP}, i}/N$, with $N=\num{100}$.

\begin{figure}
    \includegraphics[width=0.45\textwidth]{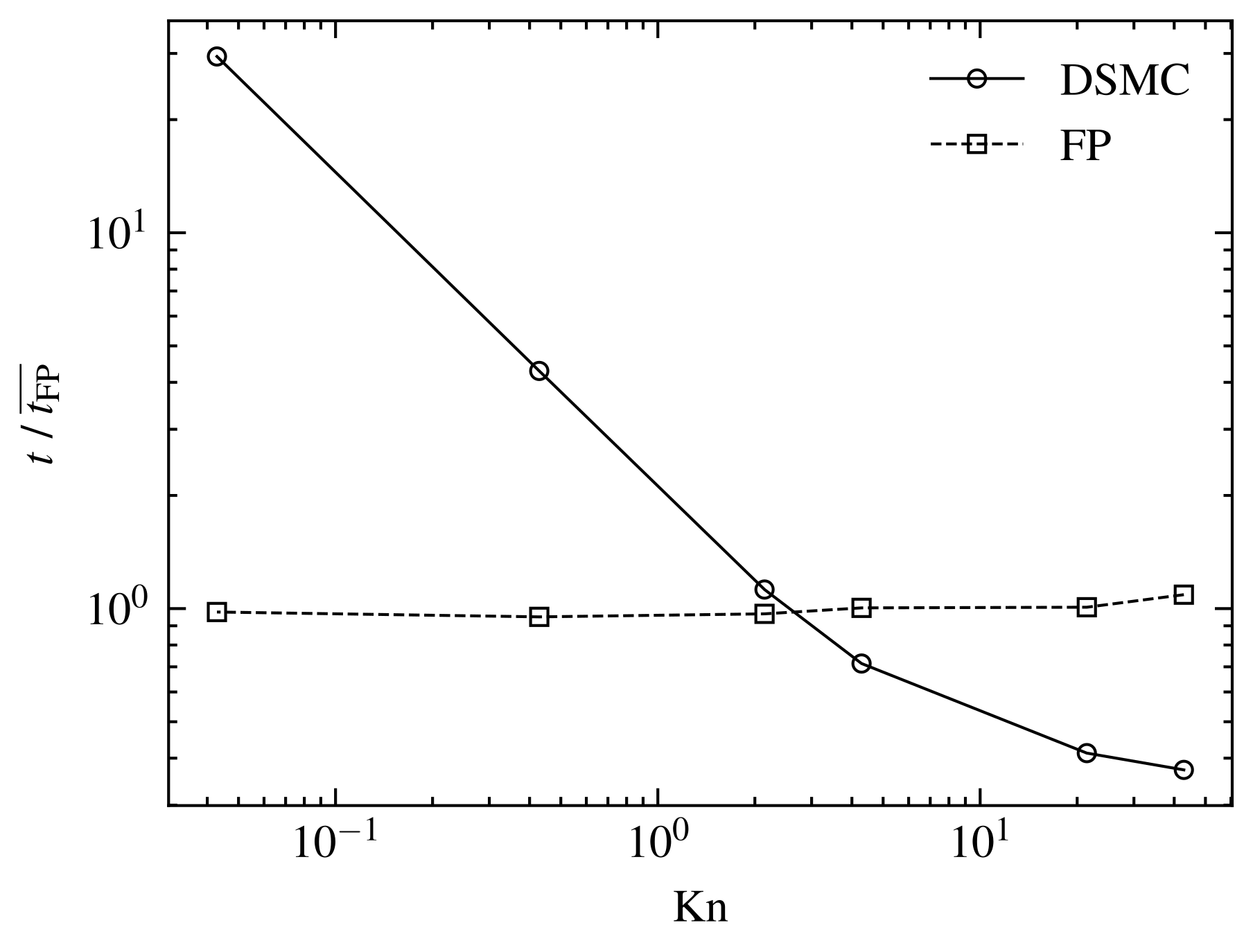}
    \caption{\label{fig:performance_FP_DSMC} Average execution time of the first \num{100} steps of the exchange reaction \ch{O2 + N -> O + NO}.}
\end{figure}

As seen in Fig.~\ref{fig:performance_FP_DSMC} the computational time rises for the DSMC simulations with decreasing Kn number while staying constant for the FP simulations.
This shows that our chemistry extension retains the performance advantage of the FP model.
For the presented test case the FP simulation becomes more efficient for Knudsen numbers of $\mm{Kn} < 2$.
\section{Modeling of Air Chemistry}
In this section, we present results for a coupled air chemistry simulation using 15 dissociation reactions from Tables \ref{tab:air_chemistry_o2_diss}, \ref{tab:air_chemistry_n2_diss}, and \ref{tab:air_chemistry_no_diss}, along with the 2 exchange reactions listed in Table~\ref{tab:air_chemistry_exchange}.
The simulation setup was chosen to be the same as described in Section \ref{sec:reaction_kinetcs_k_T}.
For this analysis, only the continuum model and the FP results were considered, as they share the closest modeling assumptions.

\begin{figure}
    \includegraphics[width=0.6\textwidth]{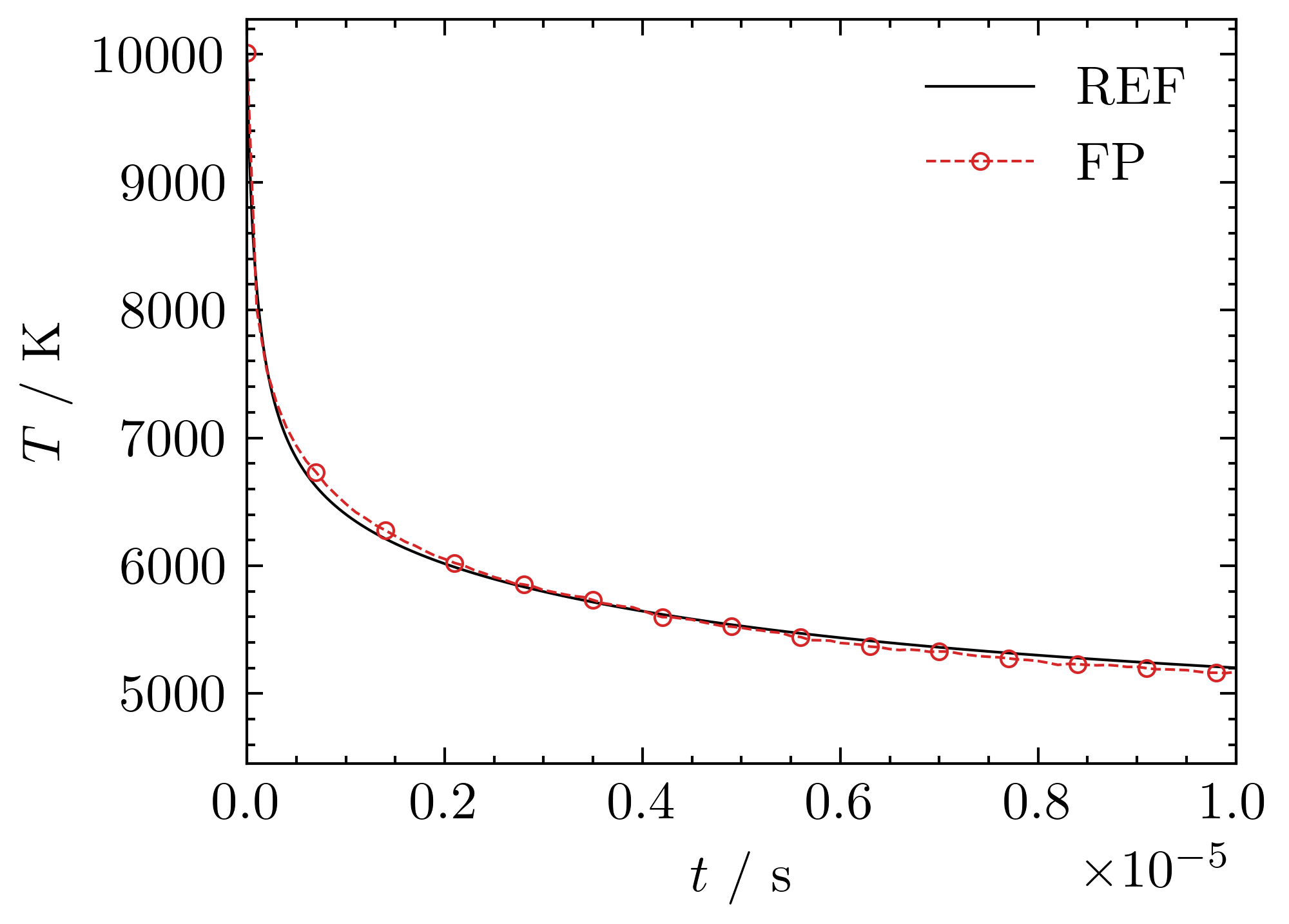}
    \caption{\label{fig:temp_air} Temperature using air reactions.}
\end{figure}

\begin{figure}
    \includegraphics[width=0.6\textwidth]{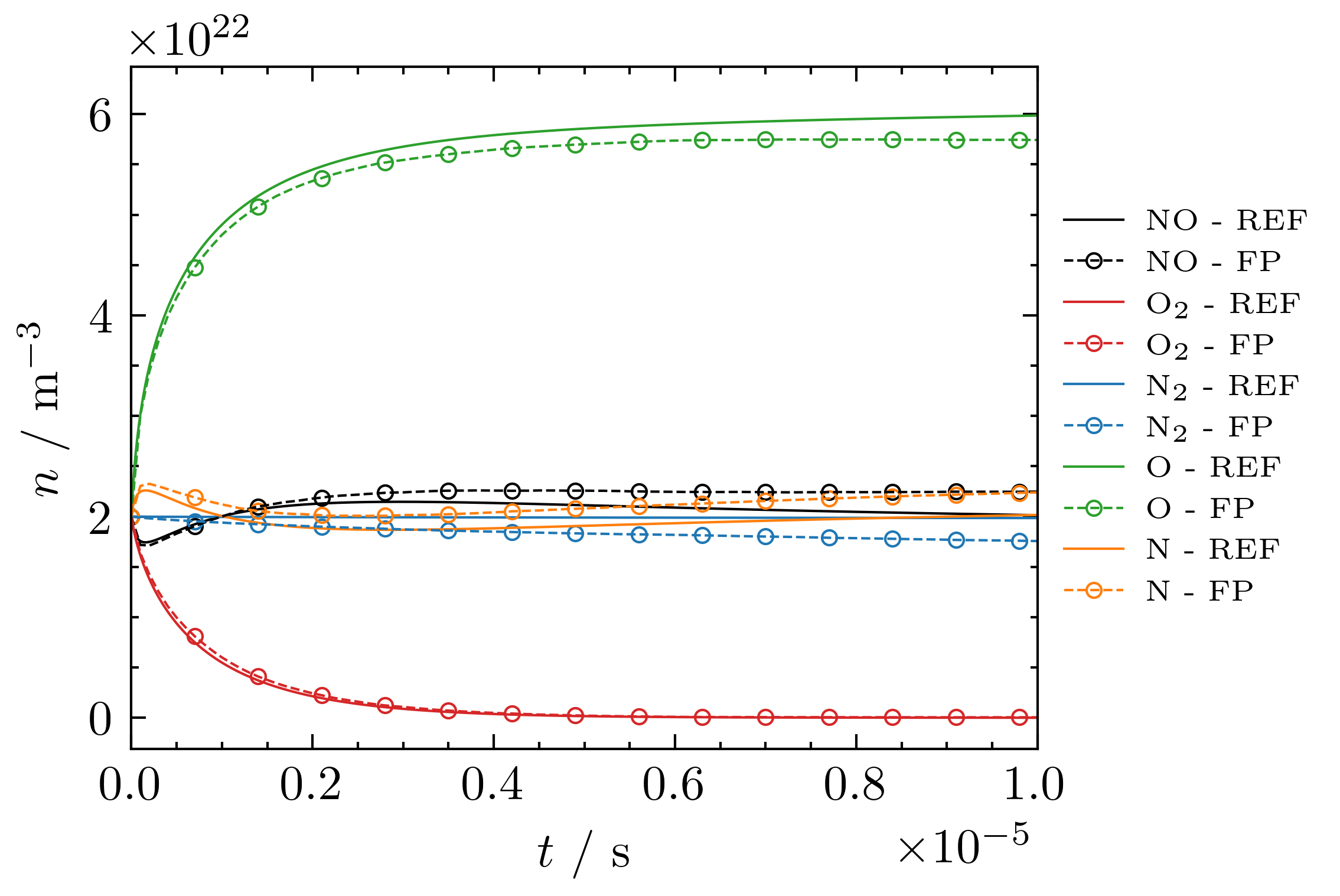}
    \caption{\label{fig:nrho_air} Number density using air reactions.}
\end{figure}

\begin{figure}
    \includegraphics[width=0.6\textwidth]{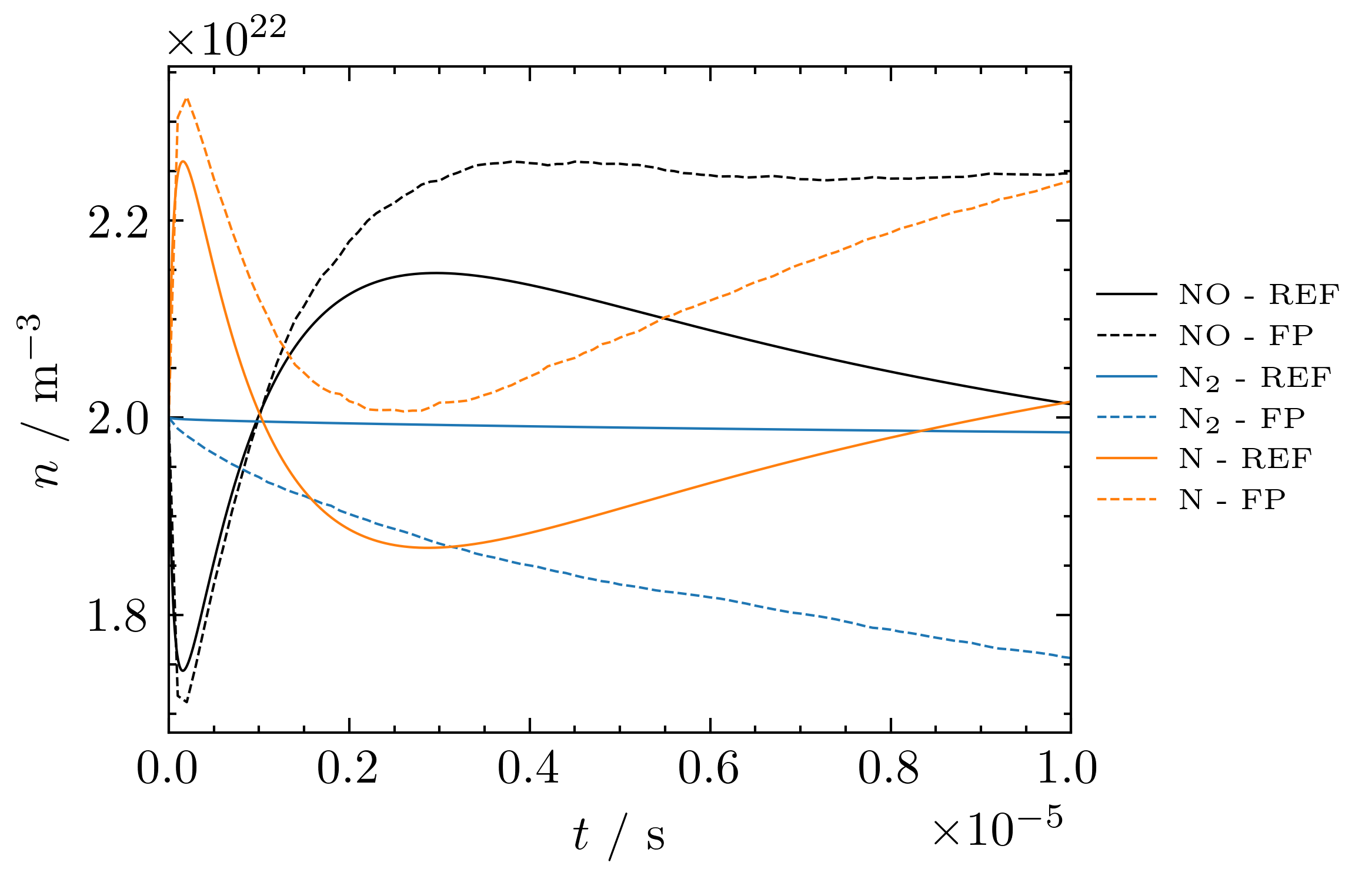}
    \caption{\label{fig:nrho_air_NO_N2_N} Number density of \ch{NO}, \ch{N2}, and \ch{N} using air reactions.}
\end{figure}

\begin{figure}
    \includegraphics[width=0.6\textwidth]{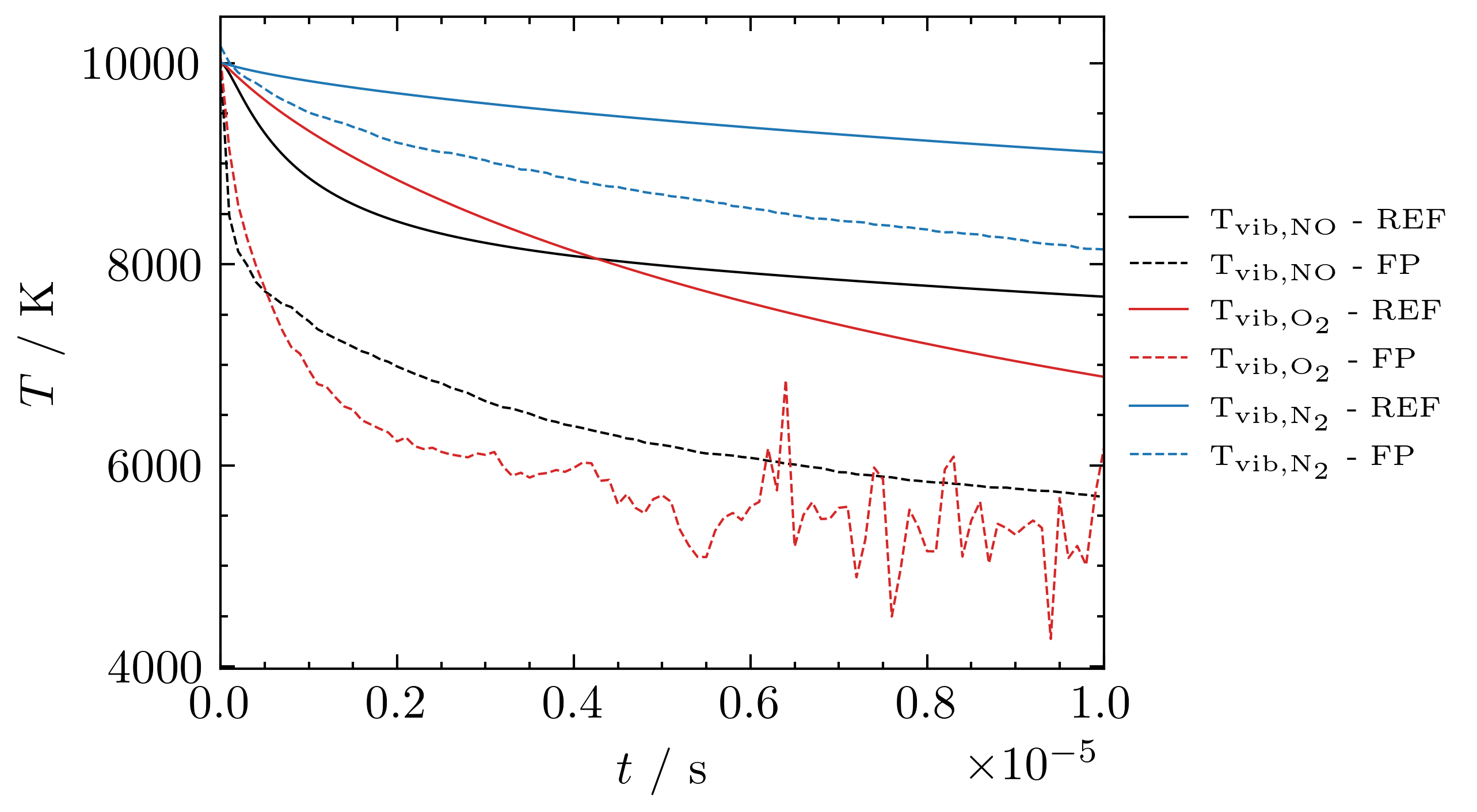}
    \caption{\label{fig:tvib_air} Temperature using air reactions.}
\end{figure}

As in previous results, the development of translational temperature agrees very well between the two models, as shown in Fig.~\ref{fig:temp_air}.
%----------------------------------------------------------------------------------------------------------------------
%----------------------------------------------------------------------------------------------------------------------
Also the development of the number densities, seen in Fig.~\ref{fig:nrho_air} shows a good agreement even in the very numerically stiff section at the beginning of the simulation.
%----------------------------------------------------------------------------------------------------------------------
%----------------------------------------------------------------------------------------------------------------------
However, the differences increase as the simulation progresses.
These differences can be better understood when only looking at the reaction kinetics of \ch{NO}, \ch{N2}, and \ch{N} as seen in Figure~\ref{fig:nrho_air_NO_N2_N}.
While the reaction kinetics of \ch{NO} and \ch{N} show a very similar qualitative behavior between the models, there is clearly a difference in the decay of \ch{N2}.
%----------------------------------------------------------------------------------------------------------------------
%----------------------------------------------------------------------------------------------------------------------
Some of the reactions involving NO proceed on very short time-scales, and the results are strongly affected by the choice of timestep and time-stepping scheme, as this rapid evolution of the system at the start of the simulation needs to be resolved.
In addition, even small differences in temperature can skew the the kinetics.
Those differences in the translational temperature can be explained by differences in the relaxation of vibrational energies as seen in Fig.~\ref{fig:tvib_air}, which, as discussed
above, is attributed to the differences between the vibrational relaxation modelling between the continuum- and particle-based approaches.
%----------------------------------------------------------------------------------------------------------------------

Nonetheless our model is able to achieve a good agreement with the continuum model even for this large and stiff system.
The observed maximum relative differences between the different methods for the number densities of all species apart from \ch{O2} do not exceed 10 - 13\%. For \ch{O2} the difference is larger (up to 100\%) due to its number density approaching zero as the simulation processes. Therefore, even minor difference in reaction rates lead to large relative changes in the number density; however, the absolute values remain negligibly small.
The maximum relative difference for the temperature is around 2\%.

\section{Conclusion and Outlook}
A framework for modeling chemical reactions within the particle-based Fokker--Planck framework has been developed.
The approach ensures the conservation of mass, momentum, and energy.
It allows for the modeling of various reaction types, such as exchange and dissociation, as well as temperature-dependent reaction rates, which are not necessarily limited to Arrhenius-based formulations.

The framework has been verified through model problems, demonstrating good agreement with solutions derived from systems of master equations as well as results produced by the SPARTA DSMC code.
Additionally, it is shown to be computationally more efficient than DSMC for moderate to low Knudsen numbers.

Planned future extensions include generalizing the framework to account for recombination reactions and the use of equilibrium rate constants for the correct modeling of backward rate coefficients.
%Further, the framework will be applied to 2- and 3-dimensional flows.
\section*{Acknowledgments}
Dr.~Georgii Oblapenko acknowledges the support of the German Science Foundation (DFG) within the research unit SFB 1481.

\appendix
\section{Chemistry Data}
%---------------------------------------------------------------------------------------------------------------
\begin{table}[h!]
\caption{\label{tab:air_chemistry_o2_diss}Air Chemistry \ch{O2} Dissociation Reactions}
\centering
 \begin{tabular}{l|ccc} 
Reaction &  $A$ / $\left(\si{\meter\cubed\per\second\kelvin\tothe{-B}}\right)$  & $B$ & $E_a$ / $\si{\joule}$\\
 \hline
\ch{O2 + N -> 2 O + N} & \num{1.660e-8} & \num{-1.5} & \num{-8.197e-19}\\
\ch{O2 + NO -> 2 O + NO} & \num{3.321e-9} & \num{-1.5} & \num{-8.197e-19}\\
\ch{O2 + N2 -> 2 O + N2} & \num{3.321e-9} & \num{-1.5} & \num{-8.197e-19}\\
\ch{O2 + O2 -> 2 O + O2} & \num{3.321e-9} & \num{-1.5} & \num{-8.197e-19}\\
\ch{O2 + O -> 3 O} & \num{1.660e-8} & \num{-1.5} & \num{-8.197e-19}\\
 \end{tabular}
\end{table}
%---------------------------------------------------------------------------------------------------------------
%---------------------------------------------------------------------------------------------------------------
\begin{table}[h!]
\caption{\label{tab:air_chemistry_n2_diss}Air Chemistry \ch{N2} Dissociation Reactions}
\centering
 \begin{tabular}{l|ccc} 
Reaction &  $A$ / $\left(\si{\meter\cubed\per\second\kelvin\tothe{-B}}\right)$  & $B$ & $E_a$ / $\si{\joule}$\\
 \hline
\ch{N2 + O -> 2 N + O} & \num{4.980e-8} & \num{-1.6} & \num{-1.561e-18}\\
\ch{N2 + O2 -> 2 N + O2} & \num{1.162e-8} & \num{-1.6} & \num{-1.561e-18}\\
\ch{N2 + NO -> 2 N + NO} & \num{1.162e-8} & \num{-1.6} & \num{-1.561e-18}\\
\ch{N2 + N2 -> 2 N + N2} & \num{1.162e-8} & \num{-1.6} & \num{-1.561e-18}\\
\ch{N2 + N -> 3 N} & \num{4.980e-8} & \num{-1.6} & \num{-1.561e-18}\\
 \end{tabular}
\end{table}
%---------------------------------------------------------------------------------------------------------------
%---------------------------------------------------------------------------------------------------------------
\begin{table}[h!]
\caption{\label{tab:air_chemistry_no_diss}Air Chemistry \ch{NO} Dissociation Reactions}
\centering
 \begin{tabular}{l|ccc} 
Reaction &  $A$ / $\left(\si{\meter\cubed\per\second\kelvin\tothe{-B}}\right)$  & $B$ & $E_a$ / $\si{\joule}$\\
 \hline
\ch{NO + N2 -> N + O + N2} & \num{8.302e-15} & \num{0} & \num{-1.043e-18}\\
\ch{NO + O2 -> N + O + O2} & \num{8.302e-15} & \num{0} & \num{-1.043e-18}\\
\ch{NO + NO -> N + O + NO} & \num{8.302e-15} & \num{0} & \num{-1.043e-18}\\
\ch{NO + O -> N + 2 O} & \num{1.862e-13} & \num{0} & \num{-1.043e-18}\\
\ch{NO + N -> 2 N + O} & \num{1.862e-13} & \num{0} & \num{-1.043e-18}\\
 \end{tabular}
\end{table}
%---------------------------------------------------------------------------------------------------------------
%---------------------------------------------------------------------------------------------------------------
\begin{table}[h!]
\caption{\label{tab:air_chemistry_exchange}Air Chemistry Exchange Reactions}
\centering
 \begin{tabular}{l|ccc} 
Reaction &  $A$ / $\left(\si{\meter\cubed\per\second\kelvin\tothe{-B}}\right)$  & $B$ & $E_a$ / $\si{\joule}$\\
 \hline
\ch{O2 + N -> NO + O} & \num{4.601e-15} & \num{-0.546} & \num{0}\\
\ch{N2 + O -> NO + N} & \num{1.069e-12} & \num{-1.0} & \num{-5.175e-19}\\
 \end{tabular}
\end{table}
%---------------------------------------------------------------------------------------------------------------

%\nocite{*}
% \bibliography{library}% Produces the bibliography via BibTeX.
\newpage

\end{document}